\documentclass[prb,a4paper,showpacs,twocolumn,superscriptaddress,longbibliography]{revtex4-1}

\usepackage{textcomp}
\usepackage{amssymb}
\usepackage{amsmath}
\usepackage{amsfonts}
\usepackage{graphicx}
\usepackage{bm}
\usepackage{xcolor}
\usepackage{multirow}
\usepackage{natbib}
\usepackage{hyperref}
\usepackage{mathrsfs}

\begin{document}

\title{Topological properties of multi-terminal superconducting nanostructures: effect of a continuous spectrum}

\author{E. V. Repin}

\affiliation{Kavli Institute of Nanoscience, Delft University of Technology, 2628 CJ Delft, The Netherlands}

\author{Y. Chen}

\affiliation{Kavli Institute of Nanoscience, Delft University of Technology, 2628 CJ Delft, The Netherlands}

\author{Y. V. Nazarov}

\affiliation{Kavli Institute of Nanoscience, Delft University of Technology, 2628 CJ Delft, The Netherlands}

\begin{abstract}
Recently, it has been shown that multi-terminal superconducting nanostructures may possess topological properties that involve Berry curvatures in the parametric space of the superconducting phases of the terminals, and associated Chern numbers that are manifested in quantized transconductances of the nanostructure.
In this Article, we investigate how the  continuous spectrum that is intrinsically present in superconductors, affects these properties. We model the nanostructure within scattering formalism deriving the action and the response function that permits a re-definition of Berry curvature for continuous spectrum.

We have found that the re-defined Berry curvature may have a non-topological phase-independent contribution that adds a non-quantized part to the transconductances. This contribution vanishes for a time-reversible scattering matrix. We have found compact expressions for the redefined Berry curvature for the cases of weak energy dependence of the scattering matrix and investigated the vicinity of Weyl singularities in the spectrum.

\end{abstract}

\maketitle

\section{Introduction}

The study of topological materials has been on the front edge of the modern research in condensed matter physics for the past decade \cite{PhysRevLett.121.087001, PhysRevLett.121.086803, PhysRevLett.121.037701, PhysRevLett.120.256601, PhysRevLett.120.130503}. These materials are appealing from fundamental point of view and for possible applications (TI-based Photodetector\cite{PhysRevB.97.081402, PhysRevApplied.8.064001}, spintronics\cite{PhysRevApplied.2.054010}, field-effect transistor\cite{PhysRevB.82.195409}, catalyst\cite{PhysRevLett.107.056804} and quantum computing\cite{RevModPhys.80.1083, PhysRevX.6.031016}). The basis for applications is the topological protection of quantum states, which makes the states robust against small perturbations and leads to many unusual phenomena, e.g. topologically protected edge states\cite{PhysRevLett.95.226801, PhysRevLett.96.106401, PhysRevLett.98.106803}. The topological superconductors\cite{PhysRevLett.102.187001, PhysRevB.82.184516, nphys2479, PhysRevLett.105.097001} and Chern insulators\cite{PhysRevLett.61.2015, PhysRevX.1.021014, PhysRevB.89.195144, PhysRevB.74.235111} are the classes of topological materials that are relevant for the present paper. In the case of the Chern insulator the topological characteristic is an integer Chern number\cite{RevModPhys.83.1057, PhysRevB.75.121306} computed with the Green's function of electrons occupying the bands in a Brillouin zone of a material - WZW form\cite{WITTEN1983422, PhysRevLett.105.256803, 1367-2630-12-6-065007,PhysRevB.84.125132}. The first Chern number reduces to the sum of first Chern numbers of the filled bands. For each band, the first Chern number is defined as an integral of the Berry curvature over the Brillouin zone\cite{PhysRevB.31.3372, PhysRevLett.49.405}. The Berry curvature is commonly defined\cite{Berry45} as $B_{\alpha \beta}=2{\rm Im} \langle \partial_\alpha k|\partial_\beta k\rangle$ with $|k\rangle$ being the wavefunction in this band and $\alpha,\beta$ being the parameters: in this case two components of a wavevector. 
If the Chern number of a crystal is not zero, the edge states necessarily appear at the interface between the crystal and the vacuum (since the Chern number of the vacuum is zero). The dimensionality of topological materials in real space is restricted by three from above, which significantly limits possible topological phases.

However, there is a way to circumvent this fundamental limitation. Recently, the multi-terminal superconducting nanostructures with conventional superconductors were proposed to realize the topological solids in higher dimensions\cite{ncomms11167}. Such nanostructures host discrete spectrum of so called Andreev bound states\cite{Andreev, DEGENNES1963151, PhysRevLett.66.3056}. The energies and wavefunctions of these states depend periodically on the phases of superconducting terminals. This sets an analogy with a bandstructure that depends periodically on the wavevectors. The dimensionality of this bandstructure is the number of terminals minus one. Also, as it was noted\cite{ncomms11167}, the multi-terminal superconducting nanostructures cannot be classified as the high-dimensional topological superconductors from the standard periodic table of topological phases\cite{doi:10.1063/1.3149495}. The authors of \cite{ncomms11167} have considered in detail 4-terminal superconducting nanostructures and proved the existence of Weyl singularities\cite{Lu622,nature15768} in the spectrum. The Weyl singularity is manifested as level crossing of Andreev bound states at a certain point in 3-dimensional phase space. Each Weyl singularity can be regarded as a point-like source of Berry curvature. Owing to this, a nonzero two-dimensional Chern number can be realized and is manifested as a quantized transconductance of the nanostructure. 
This transconductance is the response of the current in one of the terminals on the voltage applied to the other terminal in the limit of small voltage, this signifies an adiabatic regime. 

The peculiarity of the system under consideration is the presence of a continuous spectrum next to the discrete one. These states are the extended states in the terminals with energies above the superconducting gap. Were a spectrum discrete, the adiabaticity condition would imply the level spacing being much larger than the driving frequency. The level-spacing is zero for a continuous spectrum, so this complicates the adiabaticity conditions. This has been pointed out already in Ref.\cite{ncomms11167} but was not investigated in detail. We note the generality of the situation: a generic gapped system might have a continuous spectrum above the certain threshold, and the adiabaticity condition required for the manifestations of topology needs to be revisited in this situation.

The aim of the present article is to investigate this question in detail for a generic model of a superconducting nanostructure. We have studied the linear response of currents on the changes of superconducting phases in the terminals. We model a multi-terminal superconducting nanostructure within the scattering approach\cite{Transport}. In this approach the terminals of the nanostructure are described with semiclassical Green's functions and the scatterer coupled to the terminals is described by a unitary (in real time) S-matrix. Although it is not crucial, we made use of Matsubara formalism which conveniently allows us to concentrate on the ground state of the system and the limit of zero temperature is formally achieved by considering continuous Matsubara frequencies. So we do the calculations in imaginary time formalism\cite{SCHON1990237}. At the first step, we obtain the general effective action describing the nanostructure in terms of the S-matrix and time-dependent semiclassical Green's functions of the terminals. At the second step, we expand the action to the second power in time-dependent phases of the terminals. At the third step, we concentrate on the limit of small voltage and driving frequency, to obtain the response function relevant for topological properties.


We can use the properly anti-symmetrized response function as a generalized definition of the Berry curvature that is suitable for the systems with and without a continuous spectrum. The main result of the present article is that so-defined Berry curvature is contributed to by a continuous spectrum as well as discrete one even in the case of energy-independent S-matrix. We derive an explicit formula for it. This solves the paradox mentioned in \cite{ncomms11167}: the Berry curvature associated with discrete Andreev bands is discontinuous when the highest Andreev bound state merges with the continuum, which indicates that the integral of the Berry curvature defined only for discrete spectrum will not reduce to an integer. The redefined Berry curvature that we find is continuous. It gives rise to integer Chern numbers if the S-matrix is time-reversible. If it does not we reveal a specific additional non-topological contribution that does not depend on the superconducting phases. We note the the importance of the energy scales much larger than superconducting gap $|\Delta|$ in this context. This is why we also discuss in detail the case of an energy-dependent S-matrix the energy scale of variation of which may be in any relation with superconducting gap. We find that the non-topological contribution depends on the regularization of the S-matirx at large energies. In particular, it vanishes if the S-matrix is regularized as $S_{\pm \infty}=1$, this corresponds to no conduction between the terminals.

The paper is organized as follows. In Sec. \ref{Sec:system} we introduce the details of a model of a multi-terminal superconducting nanostructure and review the main aspects of a scattering matrix approach formalism in this case. The derivation and discussion of the response function are given in Sec. \ref{Sec:response}. In Sec. \ref{Sec:Weyl} we discuss the specific behaviour near the Weyl singularities, in the absence and presence of a weak spin-orbit coupling. In Sec. \ref{Sec:const} we apply the general formulae to the case of a scattering matrix that varies only slightly on the scale of the superconducting gap $|\Delta|$. In Sec. \ref{Sec:depend} we address the energy-dependent S-matrices at arbitrary energy scale for a specific model of an energy dependence. We conclude the paper with the discussion of our results (Sec. \ref{Sec:Sum}). The technical details of the derivations are presented in Appendices.

\section{Multi-terminal superconducting nanostructure \label{Sec:system}}
\begin{figure}
\centerline{\includegraphics[width=0.5\textwidth]{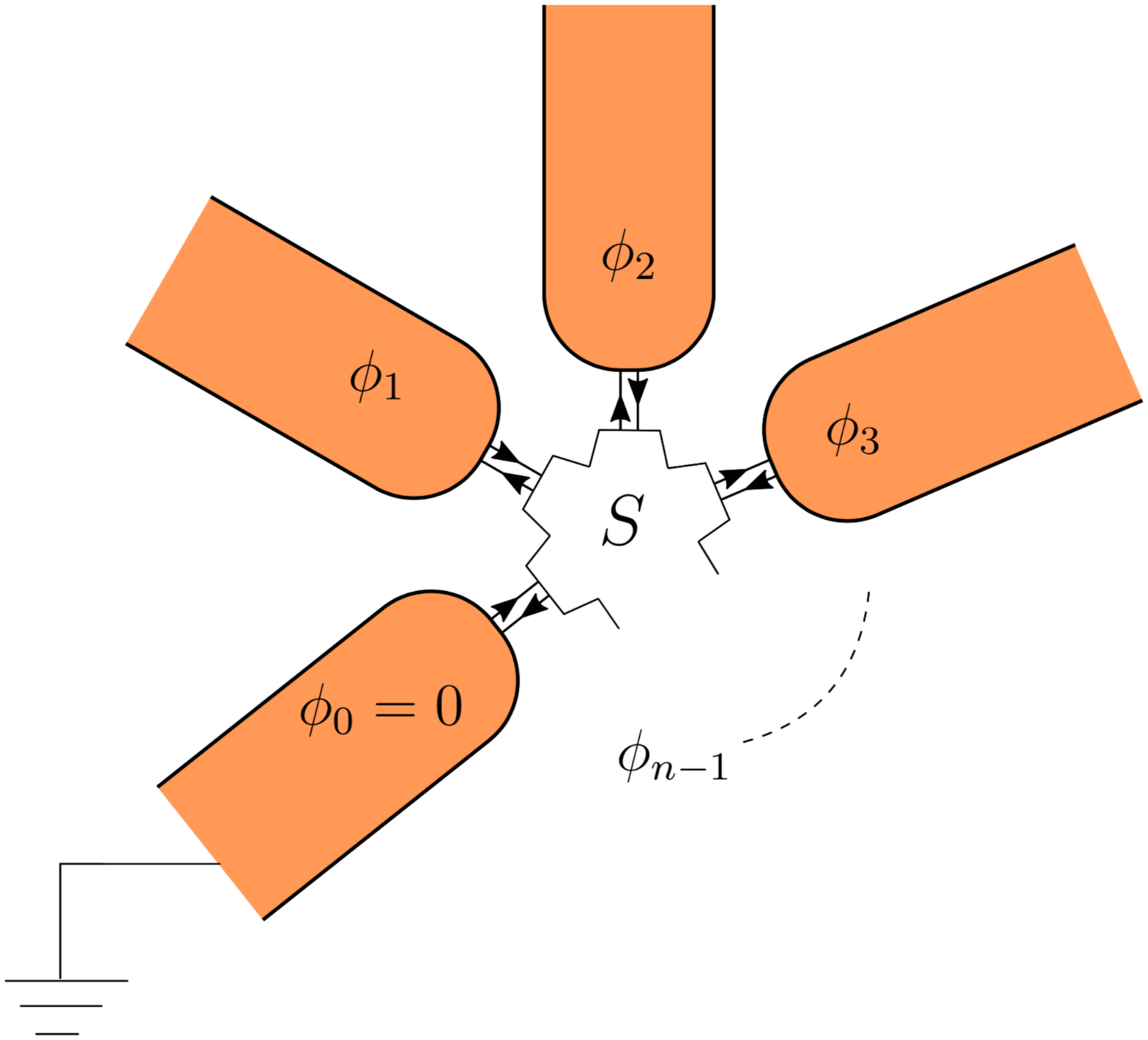}
	}
\caption{A multi-terminal superconducting nanostructure. Superconducting terminals are characterized by the corresponding superconducting phases. Electrons and holes coming from a terminal are scattered at the scattering region and can go to any other terminals. At least 4 terminals with 3 independent phases are required for a nanostructure to simulate a 3-dimensional bandstructure with topological properties.}
\label{system}
\end{figure}

Generally a multi-terminal superconducting nanostructure (Fig. \ref{system}) is a small conducting structure that connects $n$ superconducting leads. The leads are macroscopic and are characterized by the phases of the superconducting order parameter. Each lead labeled by $\alpha\in\{0,1,\cdots,n-1\}$ has its own superconducting phase $\phi_\alpha$ and one of the leads' phase can be set to zero value $\phi_0=0$, according to the overall gauge invariance. The nanostructure design and these phases determine the superconducting currents $I_\alpha$ in each lead, that are the most relevant quantities to observe experimentally. 

We aim to describe a general situation without specifying the nanostructure design. To this end, we opt to describe the system within the scattering approach pioneered by Beenakker \cite{RevModPhys.69.731}. The superconducting leads are treated as terminals: they are regarded as reservoirs which contain macroscopic amount of electrons and are in thermal equilibrium. A common assumption that we also make in this article is that all terminals are made from the same material and thus have the same modulus of the superconducting order parameter $|\Delta|$. At sufficiently low temperatures and applied voltages one can disregard possible inelastic processes in the nanostructure and concentrate on elastic scattering only. Following the basics of the scattering approach\cite{Transport}, we assume $N_\alpha$ spin-degenerate transport channels in terminal $\alpha$. The conducting structure connecting the terminals is a scattering region and is completely characterized by a scattering matrix $S$ which generally depends on energy $\varepsilon$ and is a unitary matrix at any $\varepsilon$. In Matsubara formalism we use imaginary energy $\epsilon$ and the matrix $S$ satisfies the condition $S_\epsilon S_{-\epsilon}^\dag=1$. All the details of the nanostructure design are incorporated into the scattering matrix.

The electrons and holes in the superconducting transport channels involved in the scattering process may be described as plane waves that scatter in the region of the nanostructure and then return to the corresponding terminals.  Amplitudes of incoming and outgoing waves are linearly related by the $S$-matrix. The numbers of transport channels in the terminal $\alpha$ denoted as $N_\alpha$ determines the dimension of the scattering matrix: $\textrm{dim}S = M \times M$, where $M = 2_S\sum_{\alpha} N_\alpha$ and $2_s$ counts for the spin. 
The electrons and holes experience Andreev reflection in the superconducting terminals: the electrons are converted into holes and turn back, the same happens to holes. The Andreev reflection is complete at the energies smaller than the superconducting gap $\Delta$. Therefore, electron-hole waves may be confined in the nanostructure giving rise to discrete energy levels called Andreev bound states (ABS). The amplitudes and phases of these confined states are determined by the scattering matrix and Andreev reflection phases that involve the superconducting phases of the corresponding terminals. One can find the energies of the ABS $\varepsilon$ through Beenakker's determinant equation\cite{PhysRevLett.66.3056}:
\begin{equation}
\textrm{det}(e^{2i\chi}-S_\varepsilon e^{i\phi}\sigma_y(S_{-\varepsilon}^T)^{-1}\sigma_ye^{-i\phi})=0 ,\enspace \chi = \arccos(\frac{\varepsilon}{\Delta})
\label{beenakker}
\end{equation}
where $S_\varepsilon$ is the S-matrix at the real energy $\varepsilon$, $\sigma_y=\begin{pmatrix}
0 & -i\\
i & 0
\end{pmatrix}$ is a Pauli matrix acting in the spin space and $e^{i\phi}$ is the diagonal matrix in channel space ascribing the stationary superconducting phases of the terminals to the corresponding channels, $e^{i\phi} \to \delta_{ab}e^{i\phi_\alpha}$ where $a,b$ label the channels and $\alpha$ is the terminal corresponding to the channel $a$. The ABS energies and the corresponding eigenvectors in the space of the channels depend parametrically on $n-1$ independent phases $\phi_\alpha \in [0,2\pi]$ and thus can be viewed as a bandstructure defined in a "Brilluoin zone" of phases. 
It was noted\cite{ncomms11167} that (without spin-orbit interaction) three independent parameters are needed to tune the $n-1$ dimensional band structure of energy levels of ABS to reach the Weyl singularity at zero energy. It was also noted\cite{ncomms11167} that only one parameter is required to satisfy the condition for the highest ABS to touch the continuum above the gap ($\varepsilon = |\Delta|$). The ABS merges the continuum in this case and this implies that one cannot change this level adiabatically even for arbitrarily slow change of the parameters. When the incommensurate small voltages are applied to two terminals to sweep the phases\cite{ncomms11167}, the system passes the points where the highest level merges with the continuum. This makes it questionable to apply the adiabaticity reasoning in this case. This makes it necessary to consider the contribution of the continuous spectrum to the response function of the currents in the limit of slow change of the parameters.
\section{Action\label{Sec:action}}
The most general way to describe the nanostructure under consideration is to use an action method. This method has been pioneered in the context of a simple Josephson junction in \cite{SCHON1990237}. In this method one deals with an action of the nanostructure that depends on the time-dependent superconducting phases $\phi_\alpha (\tau)$. The transport properties of the nanostructure as well as quantum fluctuations of the phases in case the nanostructure is embedded in the external circuit \cite{SCHON1990237}, can be derived from this action.

One of the advances of this Article is the derivation of such action for multi-terminal nanostructure and arbitrary S-matrix in Matsubara formalism. The details of the derivation are given in \ref{Sec:App1}. Here we give the answer:
\begin{equation}
2L=-{\rm Tr}\log[\Pi_++\Pi_-\hat{S}_\epsilon],\quad \Pi_\pm=\frac{1\pm g}{2}
\label{Bigaction}
\end{equation}
here $\Pi_\pm$ and $\hat{S}_\epsilon$ are matrices in a space that is a direct product of the space of channels, the imaginary-time space, spin and Nambu space. The matrix $\hat{S}_\epsilon$ is diagonal in the corresponding energy representation, therefore it depends on the difference of the imaginary time indices only. Its Nambu structure is given by 
\begin{equation}
\hat{S}_\epsilon=\begin{pmatrix}
S_\epsilon & 0\\
0 & S_{-\epsilon}^T
\end{pmatrix}
\end{equation}
where $S_\epsilon$ is the electron energy-dependent S-matrix (see App. \ref{Sec:App1}). The matrix $g$ is composed of the matrices diagonal in energy and diagonal in time in the following way:
\begin{equation}
g=U^\dagger \tau_z U,\quad
U^\dagger=\begin{pmatrix}
e^{\frac{i\phi(\tau)}{2}} & 0 \\
0 & e^{\frac{-i\phi(\tau)}{2}}
\end{pmatrix}
\begin{pmatrix}
A_{-\epsilon} & A_\epsilon \\
A_\epsilon & A_{-\epsilon}
\end{pmatrix}
\label{g}
\end{equation} 
where
\begin{equation}
A_\epsilon=\sqrt{\frac{E+\epsilon}{2 E}},\quad E=\sqrt{\epsilon^2+|\Delta|^2},
\label{defA}
\end{equation}
where $\tau_z$ is the 3rd Pauli matrix acting in Nambu space and the Nambu structure has been made explicit in $U^\dagger$. This form assumes that $|\Delta|$ is the same in all the terminals. If it is not so, the matrix $A_\epsilon$ also acquires the dependence on the channel index. It is worth noting that $g^2=1$ so that $\Pi_\pm$ are projectors. The matrix $g$ can be associated with the semiclassical Green's function in a terminal\cite{Eilenberger1968,Transport}: $e^{i\phi(\tau)}$ is the diagonal matrix in channel space ascribing the time-dependent superconducting phases of the terminals to the corresponding channels, $e^{i\phi(\tau)} \to \delta_{ab}e^{i\phi_\alpha(\tau)}$ where $a,b$ label the channels and $\alpha$ is the terminal corresponding to the channel $a$. 
We note the gauge invariance of the action: due to the invariance of the trace under unitary transformations, the superconducting phases can be ascribed to the terminal Green's functions $g$ as well as to the scattering matrix. 
Let us assume that the matrix $S_\epsilon$ does not depend on spin. Then the trace over spin is trivial. It is convenient to apply the unitary transformation $U^\dagger$  as in $\eqref{g}$ to all the matrices in $\eqref{Bigaction}$. This transforms the matrix $g$ to $\tau_z$. Then the projectors take a simple form $\Pi_\pm \to \frac{1\pm \sigma_z}{2}$ and the matrix in $\eqref{Bigaction}$ reduces to the lower block-triangular form in Nambu space. The determinant is then equal to the determinant of the lower right block of the transformed matrix $\bar{S}_\epsilon$. Then the action takes the form
\begin{equation}
-2L=2_S{\rm Tr}\log[A_\epsilon e^{\frac{-i\phi(\tau)}{2}}S_\epsilon e^{\frac{i\phi(\tau)}{2}}A_\epsilon+
\notag
\end{equation}
\begin{equation}
+A_{-\epsilon}e^{\frac{i\phi(\tau)}{2}}S^T_{-\epsilon}e^{\frac{-i\phi(\tau)}{2}}A_{-\epsilon}]
\label{lagrange}
\end{equation}
the S-matrix in Matsubara formalism is subject to the unitarity constraint,
\begin{equation}
\end{equation}
In what follows we concentrate on the zero-temperature limit $k_B T\ll |\Delta|$, so the summations over discrete frequencies are replaced with integrations $\int \frac{d\epsilon}{2\pi}$. 
\subsection{Stationary phases}
\begin{figure}
	\centerline{\includegraphics[width=0.5\textwidth]{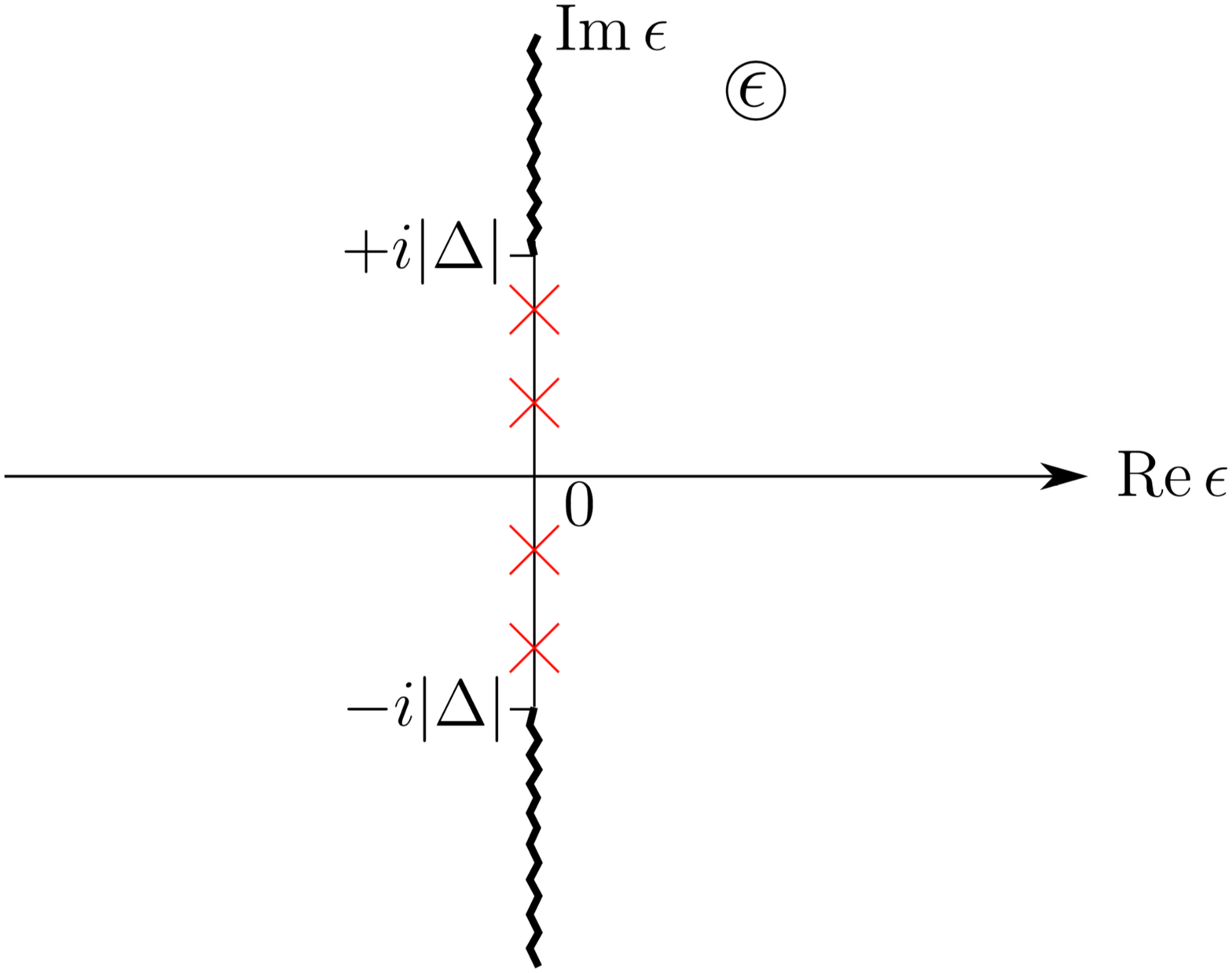}
	}
	\caption{Singularities of the matrix $Q_\epsilon$ in the complex plane of energy $\epsilon$. The symmetric cuts $[\pm i |\Delta|, \pm \infty] $ manifest the states of continuous spectrum. The isolated zeroes of the determinant of the matrix are situated at the imaginary axis within the interval $[-i|\Delta|,+i|\Delta|]$  (red crosses in the Figure). Their positions correspond to the ABS energies.}
	\label{complexplane}
\end{figure}

In the stationary case $\phi(\tau)=\phi+\delta\phi(\tau)$ with constant $\phi$ and $\delta\phi(\tau)\equiv 0$ the value of the action gives the stationary phase-dependent ground state energy of the nanostructure $E_g=\lim_{k_B T\to 0} T L_0$. 
\begin{equation}
E_g= -\frac{2_S}{2}\int \frac{d\epsilon}{2\pi} {\rm Tr} \log Q_\epsilon
\label{GSE}
\end{equation}
\begin{equation}
Q_\epsilon=A_\epsilon^2 S_\epsilon+A_{-\epsilon}^2 S_{-\epsilon}^T
\label{qmatrix}
\end{equation}
where Trace is now over the channel space and the Trace over spin space is taken explicitly as a factor of $2_S$ unless specifically addressed. The operator $Q_\epsilon$ introduced here has the properties of the inverse of the Green's function although it is not related to an operator average: its determinant as function of complex $\epsilon$ vanishes, ${\rm det} Q_{\epsilon}=0$, at imaginary values $\epsilon=\pm i \varepsilon_k$ corresponding to the ABS energies (compare with $\eqref{beenakker}$). In addition to these singularities the operator $Q_\epsilon$ has two cuts in the plane of complex $\epsilon$ corresponding to the presence of a continuous spectrum in the terminals above the gap $|\Delta|$. We choose the cuts as shown in Fig. \ref{complexplane}. The expression $\eqref{GSE}$ can be simplified in the case when the S-matrix does not depend on energy
\begin{eqnarray}
&E_g=-\frac{2_S}{2}\int \frac{d\epsilon}{2\pi}{\rm Tr}\log\left(\frac{E+\epsilon}{2E}+\frac{E-\epsilon}{2E}S S^*\right)+\\
&+\frac{2_S}{2}\int \frac{d\epsilon}{2\pi}\log\det (S^T)
\label{holycow}
\end{eqnarray}
the second (divergent) contribution here does not depend on the superconducting phases so we omit it. To compute the integral it is convenient to choose the basis in which the unitary matrix $\Lambda=SS^*$ is diagonal. This is a unitary matrix, so the eigenvalues are unimodular complex numbers. The phases of the eigenvalues are related to the energies of ABS: $\Lambda_k=e^{2i \chi_k},\quad \chi_k=\arccos[ \epsilon_k/|\Delta|],\quad\chi \in [-\pi/2;\pi/2]$. The eigenvalue $\Lambda_k=1$ is doubly degenerate and corresponds to the values $\epsilon_k=\pm |\Delta|$. The eigenvalues come in complex conjugated pairs $\Lambda_k^*=\Lambda_{-k}$, where $(-k)$ corresponds to the Nambu-counterpart of the $k-$th eigenvector. So only the eigenvalues ${\rm Im}\Lambda_k>0$ correspond to the quasiparticle states with positive energies. We will label them with positive indices $k$. In what follows we define a "bar" operation that links these pairs $|\bar{k}\rangle=S|k^\star\rangle=|-k\rangle$ where $|k\rangle$ is some eigenvector of $\Lambda$. We note, however, that this operation is not a convolution, since $|\bar{\bar{k}}\rangle=\Lambda_k|k\rangle$.

In this basis we can rewrite the integral as
\begin{equation}
E_g=-\frac{2_S}{2}\sum_{k>0}^{}\int \frac{d\epsilon}{2\pi}\log[\frac{(E+\epsilon)^2+(E-\epsilon)^2+2\cos 2 \chi_k}{4(\epsilon^2+|\Delta|^2)}]
\label{gseint}
\end{equation}
Evaluation of the integral brings to the known result
\begin{equation}
E_g=-\frac{2_S}{2}\sum_{\epsilon_k>0}^{}\epsilon_k
\label{Eg}
\end{equation}
where $\epsilon_k$ are the stationary phase-dependent ABS energies, as discussed above. The derivative of the ground state energy with respect to a stationary phase in terminal $\alpha$ gives the stationary current in the corresponding terminal,
\begin{equation}
I_\alpha=2e\frac{\partial E_g}{\partial \phi^{(0)}_\alpha}.
\end{equation}
We expect this relation to hold in the adiabatic limit. In the following Section, we will access the time-dependent currents concentrating on the next order correction in the limit of small frequencies.
\section{Response function of the currents\label{Sec:response}}
To compute the response function of the currents we assume small nonstationary phase addition to the stationary phases $\phi$, $\phi(\tau)=\phi+\delta\phi(\tau)$, $\delta\phi(\tau)\ll 2\pi$ and expand the action to the second order in $\delta \phi(\tau)$ (first order vanishes automatically since $\delta \phi(\tau)$ is nonstationary  $\int_{0}^{\beta}d\tau \delta\phi(\tau)=0$). We give the details in Append. \ref{Sec:App2}. The total contribution to the action reads 
\begin{equation}
\delta L=\sum_{\alpha,\beta}^{}\int \frac{d\omega}{2\pi}\frac{\delta\phi_\omega^\alpha \delta\phi_{-\omega}^\beta}{2}R_\omega^{\alpha \beta},
\end{equation}
$\delta \phi_\omega$ being the Fourier transform of $\delta \phi(\tau)$. The frequency-dependent response function of the current $R_\omega^{\alpha \beta}$ is given by
\begin{align}
\notag &R_\omega^{\alpha \beta}=\\
&-2_S\int \frac{d\epsilon}{2\pi} 
{\rm Tr}\Big\{Q_\epsilon^{-1}A_\epsilon^2[\frac{I_\alpha}{2}(S_{\epsilon-\omega}-S_\epsilon)\frac{I_\beta}{2}+\notag\\
&+\frac{I_\beta}{2}(S_{\epsilon+\omega}-S_\epsilon)\frac{I_\alpha}{2}]+\label{firstline}\\
&+\frac{1}{2} Q_\epsilon^{-1}\frac{\partial^2 Q_\epsilon}{\partial \alpha \partial \beta}-\label{second}\\ \notag
&-\frac{1}{2}Q^{-1}_{\epsilon+\omega}(A_{-(\epsilon+\omega)}(\frac{i I_\alpha}{2}S_{-\epsilon}^T-S_{-(\epsilon+\omega)}^T\frac{i I_\alpha}{2})A_{-\epsilon}-\\ \notag
&-A_{\epsilon+\omega}(\frac{i I_\alpha}{2}S_{\epsilon}-S_{\epsilon+\omega}\frac{i I_\alpha}{2})A_{\omega})\times\\ \notag
&\times Q^{-1}_\epsilon(A_{-\epsilon}(\frac{i I_\beta}{2}S_{-(\epsilon+\omega)}^T-S_{-\epsilon}^T\frac{i I_\beta}{2})A_{-(\epsilon+\omega)}-\\ 
&-A_{\epsilon}(\frac{i I_\beta}{2}S_{\epsilon+\omega}-S_{\epsilon}\frac{i I_\beta}{2})A_{\epsilon+\omega})\Big\}
\label{general}
\end{align}
here the stationary phases are ascribed to the S-matrix. We use a shorthand notation $\partial/\partial_\alpha=\partial/\partial_{\phi_\alpha}$ and define a set of matrices that project channel space onto the space of the channels in the terminal $\alpha$, $(I^\alpha)^{ab}=\delta^{ab}$ if $a$ is a channel in terminal $\alpha$ and $(I^\alpha)^{ab}=0$ otherwise. The first term in $\eqref{general}$ vanishes at zero frequency and in the case of the energy-independent S-matrix. The second term does not depend on frequency $\omega$. In the limit of zero frequency the second and the third terms reproduce the stationary response function of the currents

\begin{equation}
\lim\limits_{\omega \to 0} R_\omega^{\alpha \beta}=-\frac{2_S}{2}\frac{\partial^2}{\partial \alpha \partial \beta}\int \frac{d\epsilon}{2\pi}{\rm Tr}\log Q_\epsilon=\frac{\partial^2 E_g}{\partial \alpha \partial \beta}
\end{equation}
Let us consider the limit of small $\omega \ll |\Delta|$ and concentrate on the first order correction to the adiabatic limit
\begin{equation}
R_\omega^{\alpha \beta}=\frac{\partial^2 E_g}{\partial \alpha \partial \beta}+\omega B_{\alpha \beta}+O(\omega^2)
\label{linear}
\end{equation}
We note that the response function is analytic in the vicinity of $\omega=0$. This is guaranteed by the gap in the density of states,  which is given by the energy of the lowest ABS. Away from the zero-energy Weyl singularity it can be estimated as $|\Delta|/N$ with $N$ being the total number of ABS in the nanostructure. The vicinity of a Weyl singularity has to be treated more carefully as we discuss in Sec. $\ref{Sec:Weyl}$. Let us note that for any system with a discrete spectrum the quantity $B_{\alpha \beta}$ can be related to the {\it Berry curvature}\cite{Berry45, PhysRevB.31.3372, PhysRevLett.49.405}. For any state in the discrete spectrum the Berry curvature corresponding to this state is given by $B_{\alpha \beta}^{(i)}=2{\rm Im}\langle \partial_\alpha i|\partial_\beta i\rangle$ with $i$ labeling discrete states and $|i\rangle$ being the wavefunction of the corresponding state. In our case we are interested in the total Berry curvature of the superconducting ground state defined as $B_{\alpha \beta}=-\frac{1}{2}\sum_{i}^{}B_{\alpha \beta}^{(i)}$ where $i$ labels the (spin-degenerate) wavefunctions of the BdG equation with positive eigenvalues\cite{ncomms11167}. However, the adiabaticity condition which justifies the expansion in $\eqref{linear}$ for the case of discrete spectrum requires the frequency to be much smaller than the smallest energy spacing between the levels.

In our system, the continuous spectrum above the superconducting gap is present. In principle, any continuous spectrum can be approximated with a discrete spectrum with a vanishing level spacing $\overline{\delta}\to 0$. By doing this we can utilize the previous expression for the response function $B_{\alpha \beta}$ since it is valid for the discrete spectrum. However, the adiabaticity condition which is necessary for this expression to be valid would reduce to $\omega \ll \overline{\delta}\to 0$. This condition contains an artificially introduced $\overline{\delta}$ and is by construction very restrictive in $\omega$. On the other hand, the expansion in Eq. $\eqref{linear}$ is valid under a physically meaningful and less restrictive condition $\omega\ll |\Delta|/N$. Taken all that into account, we conclude that the response function $B_{\alpha \beta}$ defined in Eq $\eqref{linear}$ does not have to reduce to the expression for a total Berry curvature of a superconducting ground state of a system discussed above. The topological properties of this quantity also have to be investigated separately.

One may conjecture that the resulting response function in Eq. $\eqref{linear}$ reduces to the sum of the Berry curvatures of the discrete ABS spectrum, so that it is not contributed to by the continuous spectrum. This conjecture relies on the analogy between the expressions for the total Berry curvature and the superconducting ground state energy. In the case when the S-matrix is energy-independent, only the discrete states contribute to the ground state energy. Thus motivated, in the following we investigate the response function $B_{\alpha \beta}$ defined by means of Eq $\eqref{linear}$ in detail. We find that there is a contribution from the continuous spectrum to this quantity as well as from the discrete one. We also find that in general the integral of $B_{\alpha \beta}$ over the phases $\phi_\alpha, \phi_\beta$ that would normally define an integer Chern number, is not integer. Therefore, $B_{\alpha \beta}$ contains a non-topological contribution. This non-topological part is contributed by the continuous as well as the discrete part of the spectrum.

The tensor $B_{\alpha \beta}$ defined in Eq. $\eqref{linear}$ is antisymmetric (since $R^{\alpha \beta}_\omega=R^{\beta \alpha}_{-\omega}$). The concrete expression for $B_{\alpha \beta}$ reads:
\begin{align}
&B_{\alpha \beta}=-\frac{2_S}{2}\int \frac{{\rm d}\epsilon}{2\pi}\left(\frac{1}{2}{\rm Tr}\left[Q^{-1}_\epsilon\frac{\partial Q_\epsilon}{\partial \epsilon}Q^{-1}_\epsilon\frac{\partial Q_\epsilon}{\partial \alpha}Q^{-1}_\epsilon\frac{\partial Q_\epsilon}{\partial \beta}\right]\right.+
\notag
\\
&+
\left.\frac{\partial}{\partial \beta}{\rm Tr}\left[Q_\epsilon^{-1}A^2(\epsilon)\{\frac{\partial S_\epsilon}{\partial \epsilon},\frac{i I_\alpha}{2}\}\right]\right)-(\alpha\leftrightarrow \beta)
\label{result1}
\end{align}
The first term here resembles the usual WZW form\cite{PhysRevB.84.125132} for a Chern number. Usually, the form contains the matrix Green's functions\cite{PhysRevB.84.125132}, in our case the form utilizes the matrix $Q_\epsilon$ defined by Eq. $\eqref{qmatrix}$. We note however that in distinction from common applications of WZW forms here one cannot regard $Q_\epsilon$ as a smooth function of parameters $\phi_\alpha,\phi_\beta,\epsilon$ defined on a compact manifold without a boundary. This is because in general this matrix has different limits at positive and negative infinite energies $S_{-\infty}$ for $\epsilon \to -\infty$ and $S_{-\infty}^T$ for $\epsilon \to +\infty$ that also depend on the phases. Due to this reason the integral of the first term over a compact surface without a boundary in a space of phases does not have to reduce to an integer $\cdot(2\pi)^{-1}$. The second term in Eq. $\eqref{result1}$ has a form of a total derivative with respect to a phase of a periodic and smooth function, so the integral of this one over a compact surface will give zero.

In order to obtain the value of this integral let us consider first the variation of this value upon the small smooth variation of the matrix $Q_\epsilon \to Q_\epsilon+\delta Q_\epsilon$ that comes from the small variation of the S-matrix $\delta S_\epsilon$, so $\delta Q_\epsilon=A_\epsilon^2 \delta S_\epsilon+A_{-\epsilon}^2\delta S_{-\epsilon}^T$. The value of the integral of the second contribution in Eq. $\eqref{result1}$ does not contribute to the integral over a compact submanifold in phase space, so we needn't consider its variation. It is known~\cite{Zhang} that the variation of the first contribution to $B_{\alpha \beta}$ reduces to the total derivatives
\begin{align}
	&\delta \{\int \frac{d\epsilon}{2\pi}{\rm Tr}\left[ Q^{-1}_\epsilon\frac{\partial Q_\epsilon}{\partial \epsilon}Q^{-1}_\epsilon\frac{\partial Q_\epsilon}{\partial \alpha}Q^{-1}_\epsilon\frac{\partial Q_\epsilon}{\partial \beta}e^{\alpha \beta}\right]\}=
	\notag\\
	&=\int\frac{d\epsilon}{2\pi}\partial_\epsilon{\rm Tr}\left[Q^{-1}_\epsilon\delta Q_\epsilon Q^{-1}_\epsilon\frac{\partial Q_\epsilon}{\partial \alpha}Q^{-1}_\epsilon\frac{\partial Q_\epsilon}{\partial \beta}\right]e^{\alpha \beta}+
	\label{varcond}\\
	&+\int\frac{d\epsilon}{2\pi}\partial_\alpha{\rm Tr}\left[Q^{-1}_\epsilon\delta Q_\epsilon Q^{-1}_\epsilon(\frac{\partial Q_\epsilon}{\partial \beta}Q^{-1}_\epsilon\frac{\partial Q_\epsilon}{\partial \epsilon}-\right.\notag\\
	&\qquad\qquad \left.\frac{\partial Q_\epsilon}{\partial \epsilon}Q^{-1}_\epsilon\frac{\partial Q_\epsilon}{\partial \beta})\right]e^{\alpha \beta}
	\label{varder}
\end{align}
The value of the integral of second term in $\eqref{varder}$ over a compact submanifold in phase space vanishes if the submanifold does not pass Weyl singularities corresponding to ${\rm det}Q_{\epsilon}^{-1}\to \infty$, because it has a form of a total derivative of a smooth function. Evaluation of the integral in $\eqref{varcond}$ yields the following contribution to the variation of $B_{\alpha \beta}$
\begin{eqnarray}
\frac{1}{2\pi}\delta\{ {\rm Tr}[S_{-\infty} \frac{I_\alpha}{2}S_{+\infty}^\dagger \frac{I_\beta}{2} ]\}e^{\alpha \beta}
\label{landvar1}
\end{eqnarray}
We note that this contribution is generally nonzero and does not depend on phases. 

Let us turn to the evaluation of the topological charge that is proven to be very useful in the field~\cite{1367-2630-12-6-065007}. The value of the topological charge is defined in a usual way with the divergence of the topological field $\vec{E}$
\begin{equation}
	2\pi q={\rm div} \vec{E},\quad E^\gamma\equiv \frac{1}{2}e^{\gamma \alpha \beta} B_{\alpha \beta}
	\label{charge}
\end{equation}

To compute the topological charge we need to consider a special variation of the S-matrix that just corresponds to the stationary phase derivative $\delta S_\epsilon=[S_\epsilon,\frac{iI_\gamma}{2}]\delta \phi_\gamma$. Since the expression under the trace in $\eqref{landvar1}$ does not depend on phases, the topological charge vanishes at any point where the field $\vec{E}$ is well-defined, or alternatively ${\rm det}Q_\epsilon^{-1}$ is finite. The Weyl singularities give rise to the point-like integer charges being the sources of the field $\vec{E}$. We consider this in detail in Sec. $\ref{Sec:Weyl}$. This situation is in complete analogy with that of the standard Berry curvature of a discrete spectrum where Weyl singularities correspond to band crossings. However, we have computed the topological charge for the particular phase-dependence of the S-matrix on phases ($e^{-\frac{i\phi}{2}}S e^{-\frac{i \phi}{2}}$). We have not considered the topological charge in the space of 2 phases $\phi_\alpha,\phi_\beta$ and some other parameter characterizing the scattering matrix, this charge could be nonzero and have a continuous distribution. The investigation of the general parametric dependence of the S-matrix is beyond the scope of the present article. 

We separate the field $\vec{E}$ into three parts: a part produced by the point-like charges, divergenceless field that is zero in average, and a constant part $\vec{\bar{E}}$. The value of the integral
\begin{equation}
2\pi C^{12}=\int_{0}^{2\pi}\int_{0}^{2\pi}d\phi_1 d\phi_2 \frac{B_{\alpha \beta}e^{\alpha \beta}}{2}=\int (d\vec{s},\vec{E})
\label{flux1}
\end{equation}
is given by the flux of the topological field through the corresponding surface. This flux reduces to the integer for the first contribution to $\vec{E}$, vanishes for the second divergenceless contribution and may result in some value for the constant part of the field. We stress that the last contribution being present is the main distinction from the common case. The value of this constant field is then given by the integration of the variation $\eqref{landvar1}$:
\begin{equation}
\bar{E}^\gamma=\frac{1}{2\pi}\{ {\rm Tr}[S_{-\infty} \frac{I_\alpha}{2}S_{+\infty}^\dagger \frac{I_\beta}{2} ]\}e^{\gamma \alpha \beta}
\label{landauer}
\end{equation}
This constant field can contribute to the flux through any plane in the phase space.
\begin{equation}
C=n+2\pi(\vec{\bar{E}},\vec{n})
\label{flux}
\end{equation}
where $\vec{n}$ is the normal vector to this plane. As it has been shown in Ref.~\cite{ncomms11167} the value of $C^{12}$ is directly related to the observable transconductance between the leads $\alpha$ and $\beta$. Therefore, in contrast to the conclusions of Ref.~\cite{ncomms11167} the value of transconductance does not always quantize although the change of transconductance with a phase can be quantized.

So, in principle a nonzero non-topological contribution to $\eqref{flux}$ can be present. This contribution is nonzero if the S-matrix is not regularized at infinite energy such that $[S_{-\infty},I_\alpha]=0$. If the S-matrix is regularized in this way, then the $Q_\epsilon$ matrix is defined on a compact space of parameters $(\epsilon, \alpha, \beta)$, so the first contribution to Eq. $\eqref{result1}$ would reduce to an integer $n$ (with proper normalization). If it is not regularized this way, then this boundary term leads to the presence of a non-topological contribution to the response function, that comes due to the presence of a continuous spectrum and, formally, from the fact that the matrix $Q_\epsilon$ is not defined on a compact space, as discussed above. In the limit of energy-independent S-matrix, this contribution reduces to the antisymmetric part of the Landauer conductance\cite{Transport,5392683}. In this case, if the bare S-matrix (without the stationary phases of terminals ascribed) is non-symmetric (which means the breaking the time-reversibility condition) we obtain a nonzero value of $\eqref{landauer}$. If the S-matrix is time-reversible, the non-topological contribution is zero and the integer quantization of transconductance is restored.

\section{Weak energy dependence of the $S-$matrix\label{Sec:const}}
In the description of the realistic nanostructure a reasonable approximation is to consider the S-matrix to be constant on the scale of $|\Delta|$. It corresponds to the case of a short nanostructure (smaller than the superconducting coherence length). So a logical approximation would be to describe the nanostructure with a constant S-matrix at all energies. The response function $B_{\alpha \beta}$ is given by an integral over energy in Eq. $\eqref{result1}$. Would this integral accumulate in the region $\epsilon \sim |\Delta|$, then the approximation of a constant S-matrix at all energies would be accurate. However, there can be a significant contribution from the energy scales $\epsilon\gg |\Delta|$ to the integral yielding $B_{\alpha \beta}$. In this case the energy dependence of the S-matrix at the large energies becomes important. To investigate this we consider the contributions from the small scales $\epsilon \gtrsim |\Delta|$ and from the large scales $\epsilon \gg |\Delta|$ in the Subsections $\ref{Subsec:A}$ and $\ref{Subsec:B}$ respectively.
\subsection{Energy-independent $S-$matrix:\label{Subsec:A}}
In this Subsection we analyze the small-scale ($\epsilon\sim |\Delta|$) contribution to $\eqref{result1}$. For this we approximate the S-matrix to be constant at all energies and extend the integration limits to infinity. The second term in $\eqref{result1}$ vanishes since $\frac{\partial S_\epsilon}{\partial \epsilon}=0$. The integral in the first term in $\eqref{result1}$ converges on the scale $\epsilon \gtrsim |\Delta|$. This statement only necessarily holds if the S-matrix is energy-independent. Otherwise, the contribution from the larger scales can be present and we investigate it in $\ref{Subsec:B}$. Similarly to $\eqref{gseint}$, the result of integration under consideration can be expressed in terms of the eigenvalues and eigenvectors of the  unitary matrix $\Lambda=SS^*$. We use the same notations $\left | k \right\rangle$ and $\left | \bar{k}\right\rangle$ for the eigenvectors related to the complex conjugated eigenvalues pair $\Lambda_k$ and $\Lambda_k^*$ correspondingly as described after Eq.$\eqref{holycow}$. We remind that the phase of the eigenvalue $\Lambda_k=e^{2i\chi_k}$ with $k>0$ is related to the energy of ABS as $\chi_k=\arccos[\epsilon_k/|\Delta|]$. We also remind that $\Lambda_k=1$ is degenerate and corresponds to the energy of one of the ABS $\epsilon_k=|\Delta|$. Upon crossing this point in phase space, this ABS state exchanges the wave function with its Nambu counterpart with the eigenvalue $\epsilon_{k^\prime}=-|\Delta|$. Due to this we call such points gap touching singularities.

\begin{figure}
\centerline{\includegraphics[width=0.42\textwidth]{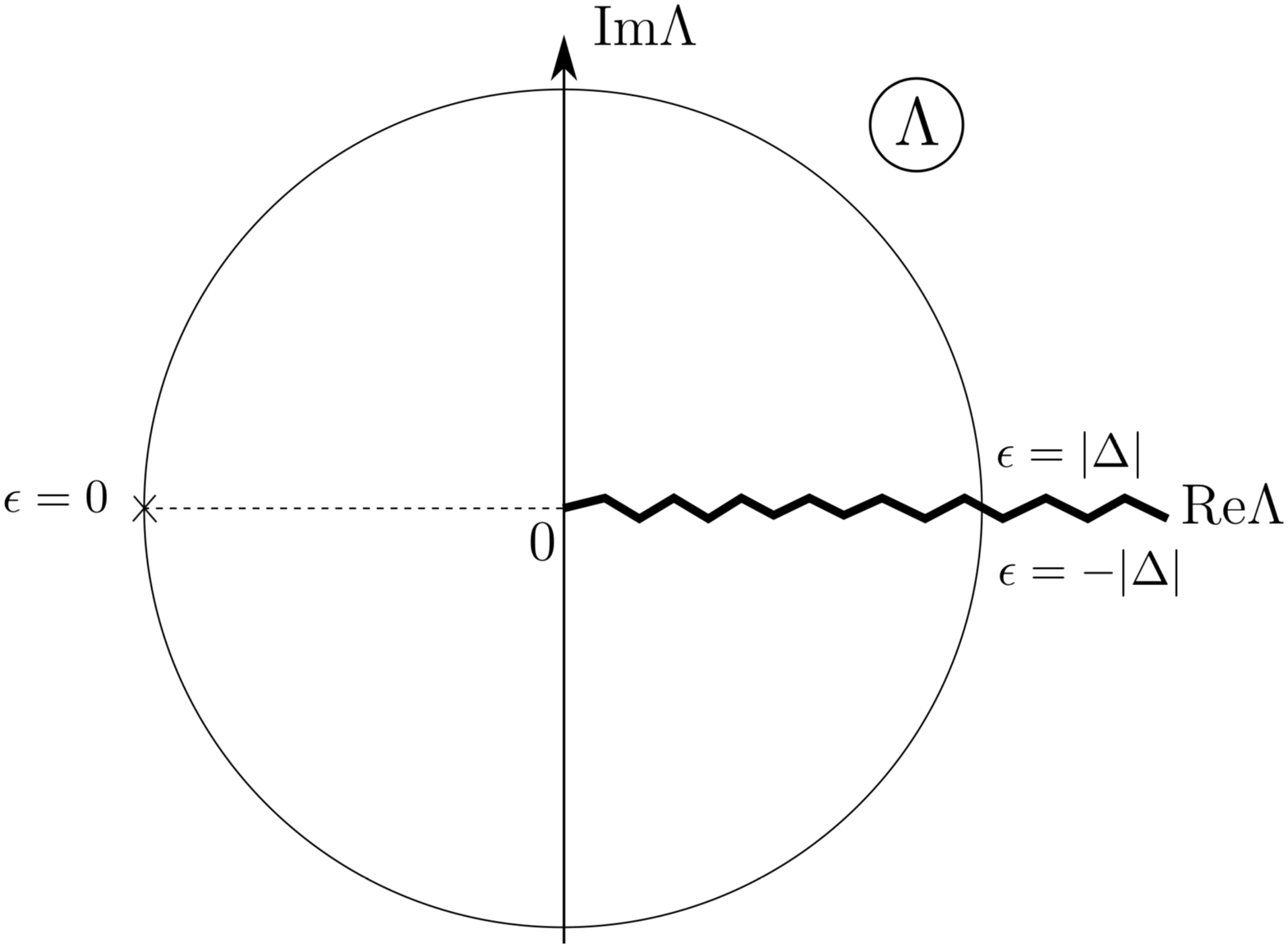}
	}
\caption{The choice of the branch cut of the logarithm in Eq. $\eqref{result2}$ in the plane of complex $\Lambda$.}
\label{branch}
\end{figure}%

\begin{figure}
	\centerline{\includegraphics[width=0.42\textwidth]{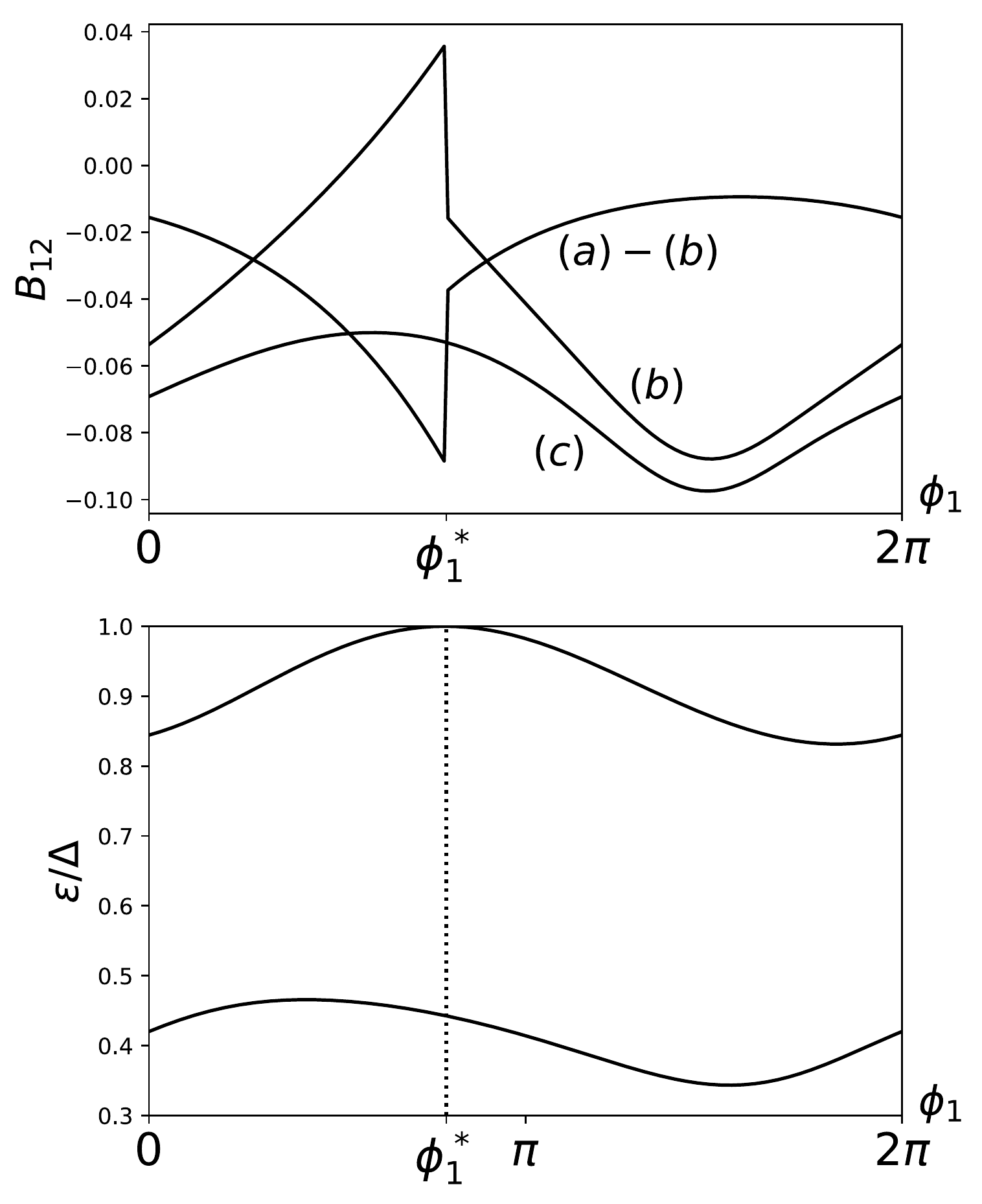}}
	\caption{ Example plots of $B_{12}$. To produce the plots, we chose one channel per terminal and took a random non-symmetric $4\times4$ scattering matrix describing the structure. We fix $\phi_2=1.20\pi, \phi_3 =0.48\pi$ and change $\phi_1$. (Upper panel) (a) the value of $B_{12}$ as given $\eqref{result2}$. It is clearly a continuous function of $\phi_1$. (b) The contribution of the discrete ABS to $B_{12}$. The  contribution experiences a jump at a point where the highest ABS merges with the continuum. (a)-(b) is thus the contribution from the continuous spectrum
	(Lower panel) The ABS energies versus $\phi_1$. The point where the highest level touches the gap egde by coincides with the point of discontinuity of the discrete spectrum contribution}
	\label{jump}
\end{figure}

Evaluating the integral yields
\begin{align}
4\pi B_{\alpha \beta}&=
-2\sum_k\big(\log\Lambda_k-\log (1+i0sgn(k))\big) \langle \partial_\alpha k|\partial_\beta k\rangle-
\notag\\
&-\sum_{k,j}^{}(1-\frac{\Lambda_k}{\Lambda_j})\langle \overline{j}|\partial_\alpha \overline{k}\rangle \langle j|\partial_\beta k\rangle-(\alpha\leftrightarrow \beta)
\label{result2}
\end{align}
where $k,j$ label the eigenvalues of $\Lambda$, and the summation goes over indices with both signs. If the number of channels is odd, there is an eigenvector of $\Lambda$ corresponding precisely to the eigenvalue $\Lambda_k=1$. Then the index $k=0$ corresponds to this state. If the number of channels is even, the indices in Eq.$\eqref{result2}$ do not take the zero value. In the following we consider the number of channels to be even. The logarithm here has a branch cut along the real axis as $[0,+\infty]$ (see Fig. \ref{branch}) to avoid the gap touching singularity ambiguity $\Lambda_k=1$. Let us consider the behaviour of $B_{\alpha \beta}$ in the vicinity of the gap touching singularity. Since the wave function corresponding to $\Lambda_k\to 1+i0$ is discontinuous upon crossing this singularity, it is not obvious that $B_{\alpha \beta}$ is continuous. However, one can observe that the first term is a sum of Berry curvatures of individual levels multiplied by the eigenvalue-dependent prefactors $\log\Lambda_k$. This prefactors vanish for the discontinuous wavefunctions at the gap touching degeneracy and guarantee the continuity of the first term. Also, one can show that the second term in Eq.$\eqref{result2}$ is continuous. Consequently, $B_{\alpha \beta}$ is continuous at this point (see Fig. \ref{jump}). The only possibility for $B_{\alpha \beta}$ to be ill-defined at some points in phase space is the zero-energy Weyl singularity where $\det Q_\epsilon^{-1}$ diverges (see Sec.$\ref{Sec:Weyl}$). 

The response function $B_{\alpha \beta}$ is expressed in terms of eigenvalues and eigenvectors of the matrix $\Lambda$. So is the ABS contribution to the ground state Berry curvature, which was conjectured as a result for $B_{\alpha \beta}$ (see Sec. $\ref{Sec:response}$). It was shown\cite{ncomms11167} that this ABS contribution is given by $B^{\rm ABS}_{\alpha \beta}=-\frac{2_S}{2}\sum_{k>0}B^{(k)}_{\alpha \beta}$, $B^{(k)}_{\alpha \beta}=2{\rm Im}\langle \partial_\alpha k|\partial_\beta k\rangle$. Since one of the wavefunctions contributing to this sum is discontinuous at the gap touching singularity, we conclude that $B^{\rm ABS}_{\alpha \beta}$ is discontinuous contrary to $B_{\alpha \beta}$. One can understand the difference between $B_{\alpha \beta}$ and $B^{\rm ABS}_{\alpha \beta}$ by considering the computation of the integral in the first term in Eq. $\eqref{result1}$ by means of complex analysis (in the plane of complex $\epsilon$). By shifting the integration contour to the upper half-plane, one can see that the integral is contributed to by the poles, corresponding to ABS and the cut above the gap (see Fig. $\ref{complexplane}$). The contribution from the poles results in $B_{\rm ABS}$, but the contribution from the cut, $B_{\alpha \beta}^{\rm cut}=B_{\alpha \beta}-B_{\alpha \beta}^{\rm ABS}\ne 0$, is equally important (see Fig. $\ref{jump}$).
\begin{figure}
\centerline{\includegraphics[width=0.42\textwidth]{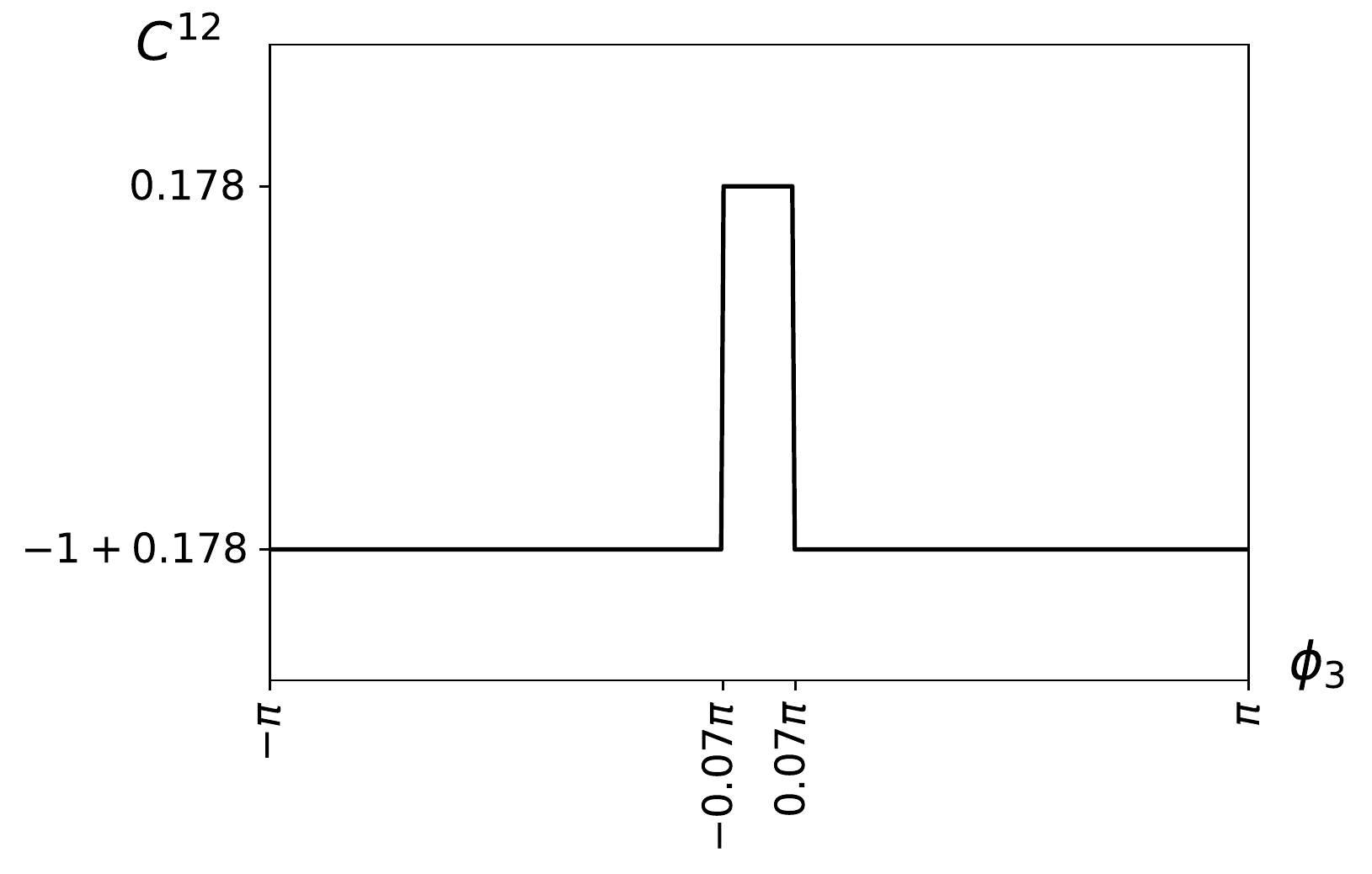}
}
\caption{An example plot of the "Chern number" $C_{12}$ defined as the integral of $B_{12}$ over $\phi_{1,2}$ (see $\eqref{flux}$). To produce the plot, we have chosen a randon $4\times4$ scattering matrix that is not invariant with respect to time reversal. We have found two Weyl singularities of opposite charge at $\phi_3 = \pm 0.07 \pi$.
We plot $C^{12}$ versus $\phi_3$ to demonstrate the integer jumps at the positions of Weyl singularities along with a non-integer, non-universal offset.
}
\label{}
\end{figure}

For the integrated $B_{\alpha \beta}$ we obtain in accordance with Eq. $\eqref{flux}$
\begin{equation}
\int_0^{2\pi}\int_0^{2\pi} d\phi_1 d\phi_2 \frac{e^{\alpha \beta} B_{\alpha \beta}}{2}= 2\pi(n+\frac{1}{4}{\rm Tr}(S^\dag I_\beta S I_\alpha)e^{\alpha \beta})
\label{Smallint1}
\end{equation}
so the value of transconductance is not necessarily quantized in the approximation of the energy-independent S-matrix.
\subsection{Contribution from the large scales$\label{Subsec:B}$}
In the previous Section we have shown that the non-topological contribution to the transconductance comes from the boundary terms at $\epsilon=\pm \infty$ (see Eq.$\eqref{varcond}$). This means that, contrary to intuition, there is an essential contribution to $B_{\alpha \beta}$ coming from the energy scales much larger than the energy gap. In order to investigate the large energy contribution we assume the regularization of the S-matrix at large energies. So, in this Subsection we consider $B_{\alpha \beta}$ for a particular energy-dependence of the S-matrix. It is chosen such that the S-matrix is regularized at infinity such that it varies slowly on the scale of a superconducting gap $|\Delta|$ and $S_{\pm \infty}=1$. This S-matrix corresponds to a complete isolation of the terminals at the largest energies. With this regularization, the matrix $Q_\epsilon$ is defined on a compact parameter space $(\alpha,\beta,\epsilon)$ and the first contribution in $\eqref{result1}$ must reduce to an integer. Due to the scale separation, there are two contributions to $B_{\alpha\beta}$. One comes from the scales $\epsilon \sim |\Delta|$ and is given by the same result $\eqref{result2}$. Another one comes from the scales $\epsilon \gg |\Delta|$. 

For negative energies, the large scale contribution with asymptotic accuracy equals
\begin{align}
&-\frac{1}{2}e^{\alpha \beta}\int_{-\infty}^{0} \frac{d\epsilon}{2\pi} {\rm Tr}[\frac{\partial S_{-\epsilon}^\dagger}{\partial \epsilon}S_{\epsilon}\frac{\partial S_{-\epsilon}^\dagger}{\partial \alpha}\frac{\partial S_{\epsilon}}{\partial \beta}]=
\notag\\
=&-\frac{1}{2}e^{\alpha \beta}\int_{-\infty}^{0} \frac{d\epsilon}{2\pi} \partial_\epsilon {\rm Tr} [S_{-\epsilon}^\dagger \frac{iI_\alpha}{2}S_\epsilon\frac{iI_\beta}{2}]=\notag\\
=&-\frac{1}{4\pi}e^{\alpha \beta}{\rm Tr}[S^\dagger \frac{iI_\alpha}{2}S \frac{iI_\beta}{2}]+\frac{1}{4\pi}e^{\alpha \beta}{\rm Tr}[S_{+\infty}^\dagger \frac{iI_\alpha}{2}S_{-\infty} \frac{iI_\beta}{2}]
\end{align}
\begin{figure}
	\centerline{\includegraphics[width=0.42\textwidth]{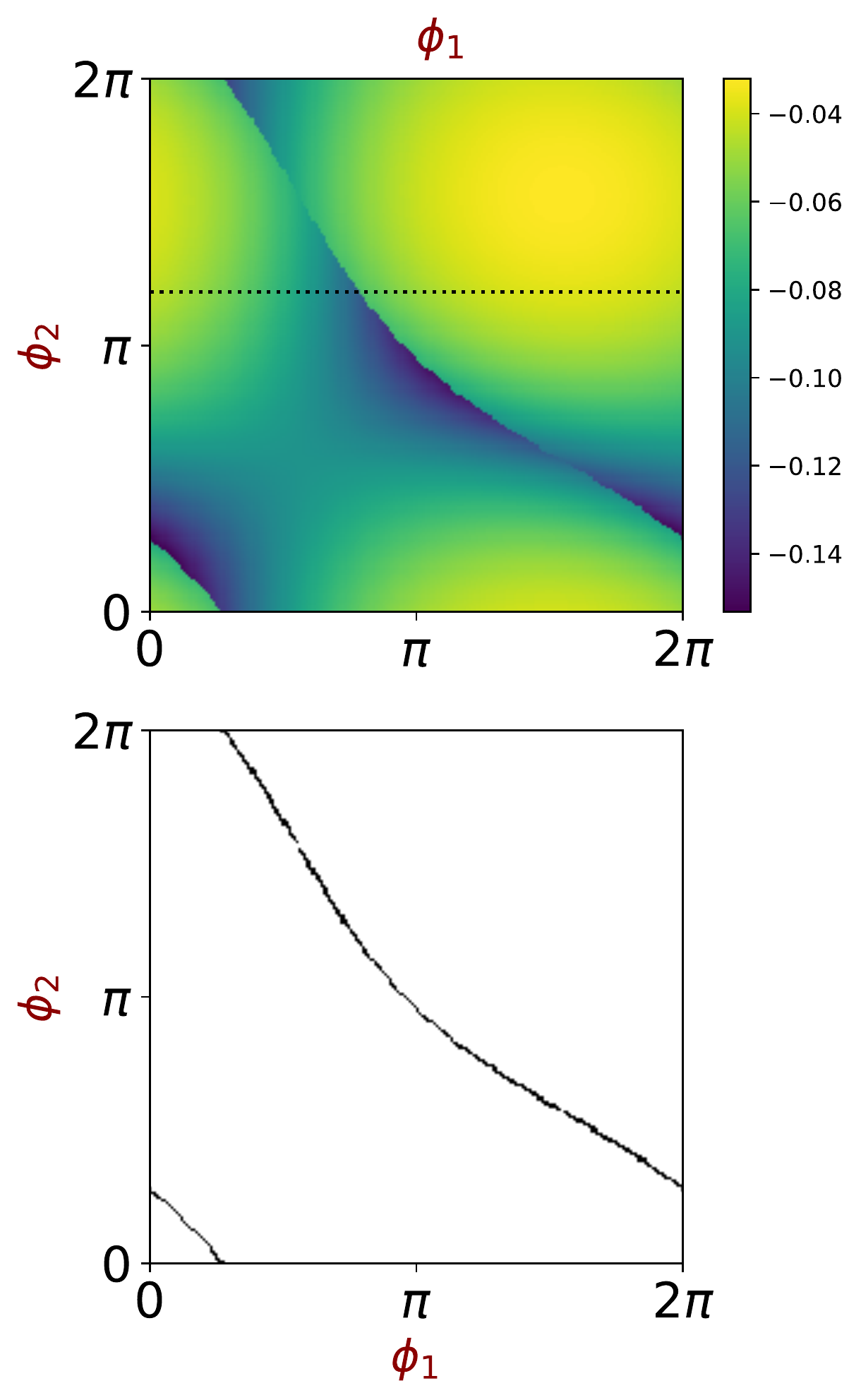}}
	\caption{Example plots versus $\phi_{1},\phi_2$. A random non-symmetric scattering matrix has been chosen to produce the plots, that varies slowly at the scale of $|\Delta|$, while $S_{\infty}=1$. Upper panel: A density plot of the continuous spectrum contribution to $B_{12}$ (\eqref{result1})versus $\phi_1,\phi_2$ at $\phi_3 = 0.48\pi$.  There is a discontinuity at the lines of the gap edge touching. Lower panel: the lines of the gap touching.}
	\label{cutcolor}
\end{figure}
with the notation $S=S_{\epsilon=0}$. 

For positive ones:
\begin{align}
&-\frac{1}{2}e^{\alpha \beta}\int_{0}^{+\infty} \frac{d\epsilon}{2\pi} {\rm Tr}[\frac{\partial S_{\epsilon}^\star}{\partial \epsilon}S^T_{-\epsilon}\frac{\partial S^\star_{\epsilon}}{\partial \alpha}\frac{\partial S^T_{-\epsilon}}{\partial \beta}]=
\notag\\
=&-\frac{1}{2}e^{\alpha \beta}\int_{0}^{+\infty} \frac{d\epsilon}{2\pi} \partial_\epsilon {\rm Tr}[S^\star_{\epsilon} \frac{iI_\alpha}{2}S^T_{-\epsilon}\frac{iI_\beta}{2}]=
\notag\\
=&-\frac{1}{4\pi}e^{\alpha \beta}{\rm Tr}[S^\dagger \frac{iI_\alpha}{2}S \frac{iI_\beta}{2}]+\frac{1}{4\pi}e^{\alpha \beta}{\rm Tr}[S_{+\infty}^\dagger \frac{iI_\alpha}{2}S_{-\infty} \frac{iI_\beta}{2}].
\end{align}
So, the both contributions give the following addition to the response function
\begin{equation}
\frac{1}{2\pi}e^{\alpha \beta}{\rm Tr}[S^\dagger \frac{I_\alpha}{2}S \frac{I_\beta}{2}]-\frac{1}{2\pi}e^{\alpha \beta}{\rm Tr}[S_{+\infty}^\dagger \frac{I_\alpha}{2}S_{-\infty} \frac{I_\beta}{2}]
\label{Large}
\end{equation}
Both terms here do not depend on phases. The first one is exactly equal to the constant part of the topological field defined previously with an opposite sign (computed for an energy-independent S-matrix case). So after integration over two phases, it cancels the non-topological contribution from small scales in $\eqref{Smallint1}$. Since we assume a regularization $S_{\pm \infty}=1$, the second term is zero (${\rm Tr}[S_{+\infty}^\dagger \frac{I_\alpha}{2}S_{-\infty} \frac{I_\beta}{2}]=0$), so the total mean value of the transconductance is quantized in correspondence with the theory of characteristic classes. 

The second contribution to $B_{\alpha \beta}$ in Eq. $\eqref{result1}$ contains the energy-derivative of the S-matrix under the integral. Due to this the energy scale of its dependence drops out from the integral. So, one may expect that it contributes to the large scale contribution to $B_{\alpha \beta}$. However, with asymptotic accuracy it vanishes in the limit when the S-matrix varies slowly on the scale $|\Delta|$. Indeed, in the limit $|\epsilon|\gg |\Delta|$
\begin{align}
\label{eq:36}
&Q_\epsilon^{-1}\simeq S_\epsilon^\star,\quad A_\epsilon^2\simeq 0,\quad \epsilon>0\\
&Q_\epsilon^{-1}\simeq S_{-\epsilon}^\dagger,\quad A_\epsilon^2\simeq 1,\quad \epsilon<0
\end{align}

In this limit for $\epsilon<0$, the integrand equals
\begin{align}
&\frac{\partial}{\partial \beta}{\rm Tr}[Q_\epsilon^{-1}A^2(\epsilon)\{\frac{\partial S_\epsilon}{\partial \epsilon},\frac{i I_\alpha}{2}\}]\simeq
\notag\\
&\simeq \partial_\beta {\rm Tr}[\frac{iI_\alpha}{2}(\frac{\partial S_\epsilon}{\partial \epsilon}S^\dagger_{-\epsilon}-\frac{\partial S_{-\epsilon}^\dagger}{\partial \epsilon}S_\epsilon)]=0
\end{align}
with asymptotic accuracy, since the expression under the trace does not depend on phases. For $\epsilon>0$ the integrand vanishes since $A_\epsilon^2\to 0$ for $\epsilon\gg |\Delta|$.


\section{The vicinity of a Weyl point} \label{Sec:Weyl}
In this Section, we investigate the Berry curvature in the vicinity of a Weyl singularity, that occurs at some point $\vec{\phi}_0$ in the 3-dimensional phase space. Such Weyl points have been analyzed in \cite{ncomms11167}  assuming spin symmetry, in \cite{PhysRevB.92.155437} the analysis has been extended to cover weak spin-orbit interaction. Without spin-orbit coupling, the Weyl points are situated at zero energy and $\det Q^{-1}_{\epsilon=0}$ diverges near the point. A conical spectrum of ABS is found in the vicinity of the point \cite{ncomms11167}. A weak spin-orbit coupling splits the energy cones  in spin and shifts the Weyl point to a finite energy \cite{PhysRevB.92.155437}. Further, we discuss separately the cases of vanishing and weak spin-orbit coupling.
\subsection{Vanishing spin-orbit coupling}
When the spin-orbit (SO) coupling is absent, the Weyl singularities are located at some points in the phase space ${\vec{\phi}_0}$ and occur at zero energy $\epsilon_{\pm} = 0$. To consider the vicinity of the singularity, we assume a small phase deviation $\delta\hat{\phi}=\hat{\phi}-\hat{\phi}_0\ll 1$ from the singularity point and assign it to each channel via the diagonal matrix $e^{\delta\hat{\phi}}$. In the vicinity, $B_{\alpha\beta}$ defined by Eq. $\eqref{result1}$ only has non-zero contributions from the first term of quasi-WZW term. The second term vanishes asymptotically when the energy approaches zero, as shown in Eq. \ref{eq:36}. Conform to these approximations, we extend the domain of the integration over the phases to infinity since $B^{\alpha\beta}$ is concentrated near the singularity point.  

To compute $B^{\alpha\beta}$, we approximate the $Q$ matrix near the Weyl point with the expression that keeps the first orders in $\epsilon$ and of the variation: $Q=(\epsilon+\frac{1}{2}\delta\Lambda)S^T = MS^T$, $S$ being the scattering matrix in the singularity point at $\epsilon=0$. Conveniently, we can replace $Q$ with $M$ in Eq.\eqref{result1}. We find the variation $\delta\Lambda$ by expanding the S-matrix in $\delta\vec{\phi}$: 
\begingroup
\allowdisplaybreaks
\begin{align}
&S\rightarrow S+\delta_\phi S=e^{-i\delta\hat{\phi}/2}Se^{i\delta\hat{\phi}/2}=S-[\frac{i\delta\hat{\phi}}{2},S]\\
&\Lambda=SS^*\rightarrow\Lambda+\delta_\phi\Lambda=\Lambda+iS\delta\hat{\phi }S^\dag\Lambda-i\delta\hat{\phi}\Lambda
\end{align}
\endgroup%

We can contract the dimension of $M$ projecting it to two eigenvectors of $\Lambda$ that achieve singular values at  the Weyl point. Following \cite{ncomms11167}, we separate the singular part of $M$ and write in the basis of ABS eigenvectors $\left | + \right\rangle$ and $\left | - \right\rangle$ satisfying $S\left | \pm \right\rangle = \pm \left | \mp \right\rangle^*$, $\Lambda |\pm\rangle = -|\pm\rangle$:
\begin{equation}
M=\epsilon+\frac{1}{2}\delta\Lambda \equiv \epsilon +\frac{i}{2}\vec{h}\cdot\vec{\tau}
\label{QnoSO}
\end{equation}%
where $\vec{\tau}$ are the Pauli matrices in the space of these two eigenvectors, and the components of $\vec{h}$ are proportional to the components of $\vec{\phi}$: $h_x+ih_y=2\left\langle - \right |\delta\hat{\phi}\left | + \right\rangle$, $h_z=\left\langle + \right |\delta\hat{\phi}\left | + \right\rangle-\left\langle - \right |\delta\hat{\phi}\left | - \right\rangle$.

The form of $M$ is similar to the generic form of Green's function of a two-level system. We expect that the two  poles of $M^{-1}$ should be positioned symmetrically on the imaginary axis $\epsilon$ due to BdG particle-hole symmetry. Indeed, we find these poles at $\epsilon_\pm = \pm i\frac{|\vec{h}|}{2}$. Using the trace relations of Pauli matrices, we reduce in the leading order $B_{\alpha\beta}$ to the Berry curvature of the corresponding levels :
\begingroup
\allowdisplaybreaks
\begin{align}
&B^{\alpha\beta}=-\frac{1}{4}\int \frac{\mathrm{d}\epsilon}{2\pi} {\rm Tr}\left(M^{-1}_\epsilon\frac{\partial M_\epsilon}{\partial \epsilon}M^{-1}_\epsilon\frac{\partial M_\epsilon}{\partial \alpha}M^{-1}_\epsilon\frac{\partial M_\epsilon}{\partial \beta}\right) \nonumber\\
&=\frac{1}{8}\int\frac{\mathrm{d}\epsilon}{2\pi}\sum_{\substack{a,b,c\\=x,y,z}}\frac{1}{(\det M)^2}\Big(h_a\partial_\alpha h_b\partial_\beta h_c \epsilon_{abc}-\nonumber\\
&-(\alpha\leftrightarrow\beta)\Big) =\frac{\vec{h}}{4|\vec{h}|^3} \cdot\partial_\alpha \vec{h} \times \partial_\beta \vec{h}-(\alpha\leftrightarrow\beta)
\end{align}
\endgroup%

We note that in this section all the matrices have the spin index. For an $N$ dimensional space of superconducting phases,  the singularities are concentrated in the $N-3$ dimensions and the relevant space is reduced to a $3$-dimensional subspace $\{\delta\phi_1,\delta\phi_2,\delta\phi_3\}$. For certainty,  we set the indices $\alpha,\beta = 1,2$, and consider the curvature defined in the $\phi_1-\phi_2$ plane at a fixed phase $\phi_3$. 

The $\phi_3$ dependence of the integral of the curvature with respect to superconducting phases $\phi_1$, $\phi_2$ witnesses the change of first Chern number $C^{12}$ when the integration plane passes the singularity point. Since we only concentrate on the vicinity of the Weyl singularity, the integral under the approximations made can only indicate the change of the Chern number, rather than its total value that can be determined by integration over the regions far from the singularity point.  To compute the integrated $B_{\alpha\beta}$, we notice from Eq.\eqref{QnoSO} that the energy spectrum is linear in $\delta\phi$, and introduce a linear relation $h_i=\sum_\alpha\delta\phi_\alpha T_{\alpha i}$ with $T_{\alpha i}=\partial_\alpha h_i$ being a real invertible matrix. The integrated $B_{12}$ is then obtained as:
\begingroup
\allowdisplaybreaks
\begin{align}
C^{12}=\frac{1}{2\pi}\int B_{12}\mathrm{d}\phi_1\mathrm{d}\phi_2= \frac{1}{2}\textrm{sgn}(\delta\phi_3\det T)
\label{Chern}
\end{align}
\endgroup%
$\textrm{sgn}(\delta\phi_3)$ determining the orientation of the $\delta\phi_3$ deviation. 

This implies that whenever the integration plane  passes the Weyl point, the first Chern number is changed by $\Delta C^{12} = \frac{1}{2}\textrm{sgn}(\delta\phi_3\det T)-\frac{1}{2}\textrm{sgn}(-\delta\phi_3\det T) =\pm 1$. This manifest the the integer values of the topological charge. The integrated $B_{\alpha\beta}$ in Eq.\eqref{Chern} specifies the flux of the Berry field penetrating the plane which is either above or below the singularity point. This flux, owing to symmetry, is a half of the total flux,  this explains the half-integer values. Therefore, the main contribution to Eq.$\eqref{result2}$ in the vicinity the Weyl point is given by the Berry curvatures of the two levels that are close to zero energy, and can be presented as
\begin{equation}
2\pi B_{\alpha \beta}=2\pi i [\left\langle \partial_\alpha + \right | \left.\partial_\beta + \right\rangle-\left \langle \partial_\alpha- \right |\left.\partial_\beta -\right\rangle]
\end{equation}
\subsection{Weak Spin-Orbit Coupling}
Let us turn on a weak spin-orbit interaction and take it into account perturbatively giving a small spin-dependent change to  the scattering matrix that preserves its unitarity, as is done in \cite{PhysRevB.92.155437}. 
The first order variation thus reads
\begingroup
\allowdisplaybreaks
\begin{align}
S\rightarrow & e^{-i\delta\phi/2}Se^{i\vec{\sigma}\cdot\vec{K}}e^{i\delta\phi/2} \nonumber\\
= &S+\delta_\phi S+iS(\vec{\sigma}\cdot\vec{K})\\
\Lambda =&S\sigma_yS^*\sigma_y\rightarrow\Lambda+\delta_\phi\Lambda+ \delta_K\Lambda\nonumber\\
=&\Lambda+\delta_\phi\Lambda+iS(\vec{\sigma}\cdot\vec{K})S^\dag\Lambda+i\Lambda(\vec{\sigma}\cdot\vec{K}^*)\end{align}
\endgroup%
where the last equality sign implies the commutation relation $\sigma_y\sigma_i^*\sigma_y=-\sigma_i$. Here,  $\vec{\sigma}$ are the Pauli matrices in spin space and $\vec{K}$ being the corresponding Hermitian matrix in the channel space characterizing the spin-orbit effects. Owing to the time reversibility, $\vec{K}(\vec{\phi}) = - \vec{K}(-\vec{\phi})$, yet in the vicinity of the singularity we may disregard its dependence on superconducting phases. 

As in the previous Subsection, we project the matrix $Q$ onto singular subspace that has now dimension of $4$ to account for spin, and replace it with the matrix $M$. Writing the latter in the basis of eigenvectors $\left | \pm \right\rangle\left | \uparrow(\downarrow)\right\rangle$:
\begingroup
\allowdisplaybreaks
\begin{align}
M=\epsilon+\frac{1}{2}\delta\Lambda=\epsilon+\frac{i}{2}(\vec{h}\cdot\vec{\tau} - \vec{\sigma}\cdot \vec{K'})
\end{align}
\endgroup%
$\vec{K}' = \left\langle + \right | \vec{K}^*\left | + \right\rangle + \left\langle - \right | \vec{K}^*\left | - \right\rangle$.  We can conveniently choose the spin quantization axis in the direction of $\vec{K}'$ replacing the operator $\vec{\sigma}\cdot \vec{K'}$ with its eigenvalues $\pm|K_0| = \pm\sqrt{|\vec{\sigma}\cdot\vec{K}'|}$ for spin up and down, respectively.

\begin{figure}
\begin{center}
\includegraphics[width=0.5\textwidth]{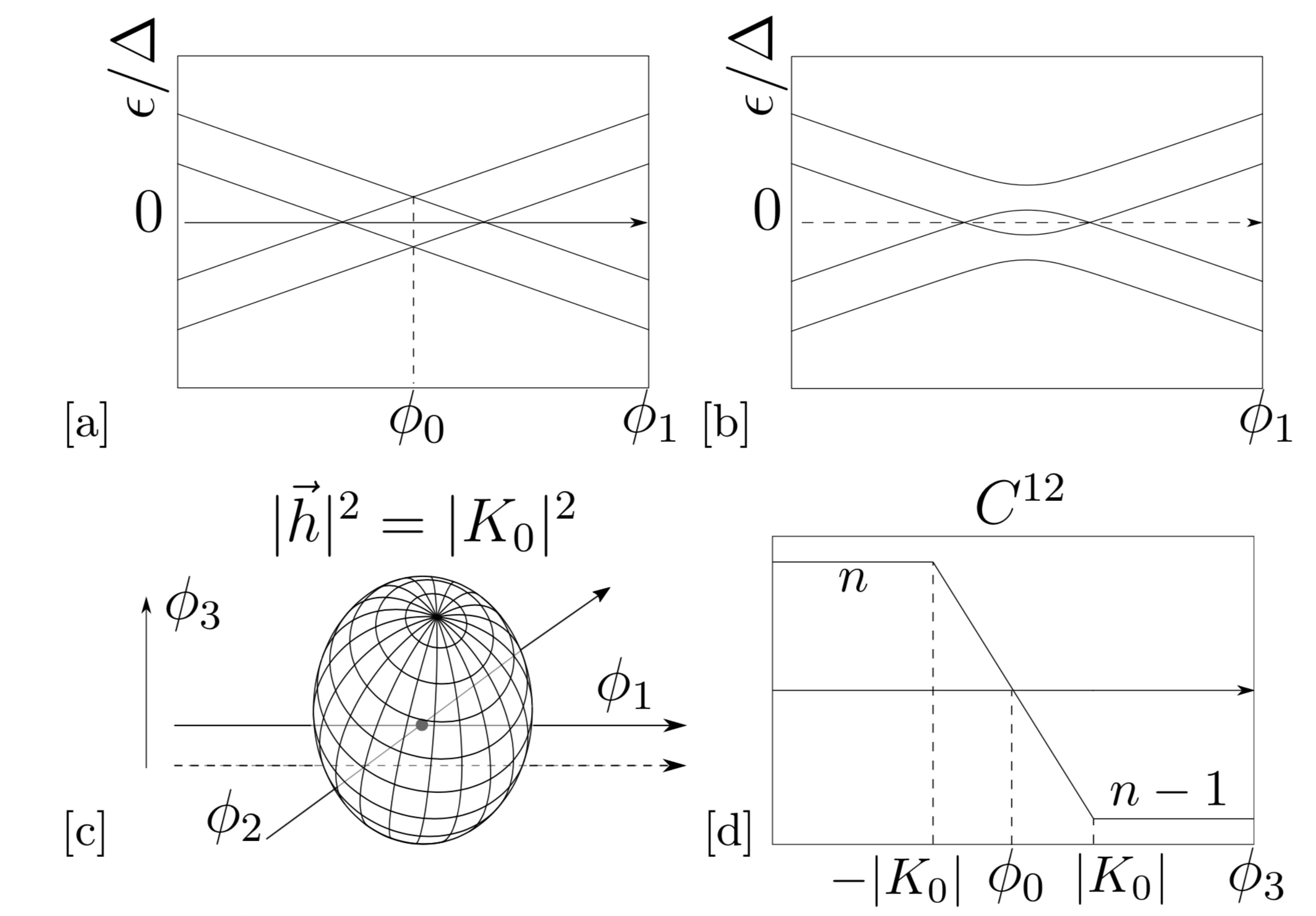}
\caption{Spin-orbit splitting of Weyl singularity [a]: ABS energies versus $\phi_1$ through the singularity for a choice $\phi_{2,3}$ corresponding to the singularity. The cone shifted upward(downward) specifies spin up (down).  [b]: ABS energy with the same $\phi_{2,3}$ along the line $\phi_1$ that misses the singularity. [c]: The ABS cross zero energy at the surface of the ellipsoid depicted. The ellipsoid encloses the singularity (central point). The ground state within the ellipsoid is of odd parity and the Berry curvature is zero. The central dot is the the Weyl singularity $\phi_0$ enclosed in the ellipsoid. The ABS energies in [a,b] are plotted along the solid [a] and dashed [b]lines in the Figure. [d]: The "Chern number"$C_{12}$ versus $\phi_3$. The topological quantization is absent owing to the discontinuity of the ground state at the surface of the ellipsoid.}
\label{levelSO}
\end{center}
\end{figure}

The spin-orbit coupling lifts the spin degeneracy of the ABS in the vicinity of a Weyl point. The poles at imaginary energies become  $\epsilon_{\uparrow} = i(\pm\frac{|\vec{h}|}{2} + \frac{|K_0|}{2})$ for spin up and $\epsilon_{\downarrow} = i(\pm\frac{|\vec{h}|}{2} - \frac{|K_0|}{2})$ for spin down. Contrary to the spin-degenerate case, the singularities at $|\vec{h}|=0$ are no longer at zero energy. Instead, they are shifted to $\pm i|K_0|$, see Fig. \ref{levelSO}. The conical  singularity of the spectrum  remains and the topology is still protected, as we will explain below in detail.

The ABS energies cross zero energy when 
\begin{equation}
|K_0| = |\vec{h}|=\sqrt{\sum\delta\phi_\alpha X_{\alpha\beta}\delta\phi_\beta}
\label{ellip}
\end{equation} 
is satisfied. (Here, we introduce a positively defined matrix $X_{\alpha\beta}=\sum_i T_{\alpha i} T_{i\beta}$. Eq.\eqref{ellip} defines an ellipsoidal surface in the 3D superconducting phase space that encloses the singularity at $\hat{\phi}_0$ where $|\vec{h}|=0$. Outside the ellipsoid, two positive imaginary poles at $\epsilon_{+\uparrow(\downarrow)} = \frac{i}{2}(|\vec{h}| \pm |K_0|)$ hold a half of the residue of the spin degenerate pole $\epsilon_+$ each. Two negative imaginary poles $\epsilon_{-\uparrow(\downarrow)}$ at $\epsilon_{-\uparrow(\downarrow)} = \frac{i}{2}(-|\vec{h}| \pm |K_0|)$ have the opposite residues. Inside the ellipsoid, poles of $\epsilon_{+\uparrow}$ and $\epsilon_{-\downarrow}$ exchange their values as well as wave functions, thus canceling the contributions from the other two poles. Thus, $B_{\alpha\beta}$ is zero inside the ellipsoid and is the same as in the spin-degenerate case outside the ellipsoid,
\begingroup
\newcommand\numberthis{\addtocounter{equation}{1}\tag{\theequation}}
\allowdisplaybreaks
\begin{align}
B^{\alpha\beta}=&\left\{
\begin{array}{lr}
\frac{\vec{h}}{4|\vec{h}|^3}\cdot\partial_\alpha \vec{h} \times \partial_\beta \vec{h}-[\alpha\leftrightarrow\beta\nonumber],&   |K_0| < |\vec{h}| \\
0, &   |K_0| > |\vec{h}|
\end{array}
\right.\numberthis
\label{BerrySO}
\end{align}
\endgroup%

The result of integration of $B^{12}$  over two superconducting phases $\phi_1$,$\phi_2$ at a fixed $\delta\phi_3$ thus reads 
\begin{equation}
C^{12} = \frac{1}{2\pi}\int \mathrm{d}\phi_1\mathrm{d}\phi_2 B^{12}\theta(|\vec{h}|\geq|K_0|^2)
\label{chern}
\end{equation}
One can understand this result geometrically by presenting Eq. \eqref{chern} as an integral 
over the corresponding plane in $\vec{h}$ space,
\begingroup
\allowdisplaybreaks
\begin{align}
C^{12} &= \frac{1}{2\pi}\int_{|\vec{h}^2|>|K_0|^2} \Big(\frac{\vec{h}}{2|\vec{h}|^3} \cdot \hat{n}_{12}^h \Big)\textrm{d}^2h_{12} \nonumber\\
&=\frac{\textrm{sgn} (\delta \phi_3\det T)}{4\pi}\int_{|\vec{h}^2|>|K_0|^2} \frac{\textrm{d}^2h_{12}}{h^2}\notag\\
&=\frac{\textrm{sgn} (\delta \phi_3\det T)}{2}\frac{\Omega_{12}}{2\pi}
\end{align}
\endgroup%
where $\hat{n}_{12}^h$ is the vector normal of the corresponding plane and $\Omega_{12}$ is eventually the solid angle at which a part of the $\phi_1-\phi_2$ plane outside the ellipsoid is seen from the Weyl singularity (see Fig.\ref{levelSO}). Generally, this angle is expressed through elliptic integrals.

The integral can be simplified if we choose the coordinate system in 3D space of the phases in such a way that $T_{13}=T_{31}=T_{23}=T_{32}=0$. With this, the integral can be evaluated as
\begingroup
\allowdisplaybreaks
\begin{align}
C^{12} 
&=\frac{\textrm{sgn}(\det T)\delta\phi_3}{2}\int\displaylimits_1^\infty\frac{(|K_0|^2-T_{33}^2\delta\phi_3^2)r\mathrm{d}r}{[(|K_0|^2-T_{33}^2\delta\phi_3^2)r^2+T_{33}\delta\phi_3^2]^{\frac{3}{2}}}\nonumber\\
&= \frac{1}{2}\textrm{sgn}(\det T)\frac{\delta\phi_3}{|K_0|}
\end{align}
\endgroup 

We see that in the vicinity of a Weyl point the $C^{12}$ is not a topologically protected quantity confined to the integer values: rather, it changes linearly in an interval of $\delta \phi_3$ defined by the strength of the spin-orbit coupling (Fig. \ref{levelSO} )

\begin{figure}
\begin{center}
\includegraphics[width=0.45\textwidth]{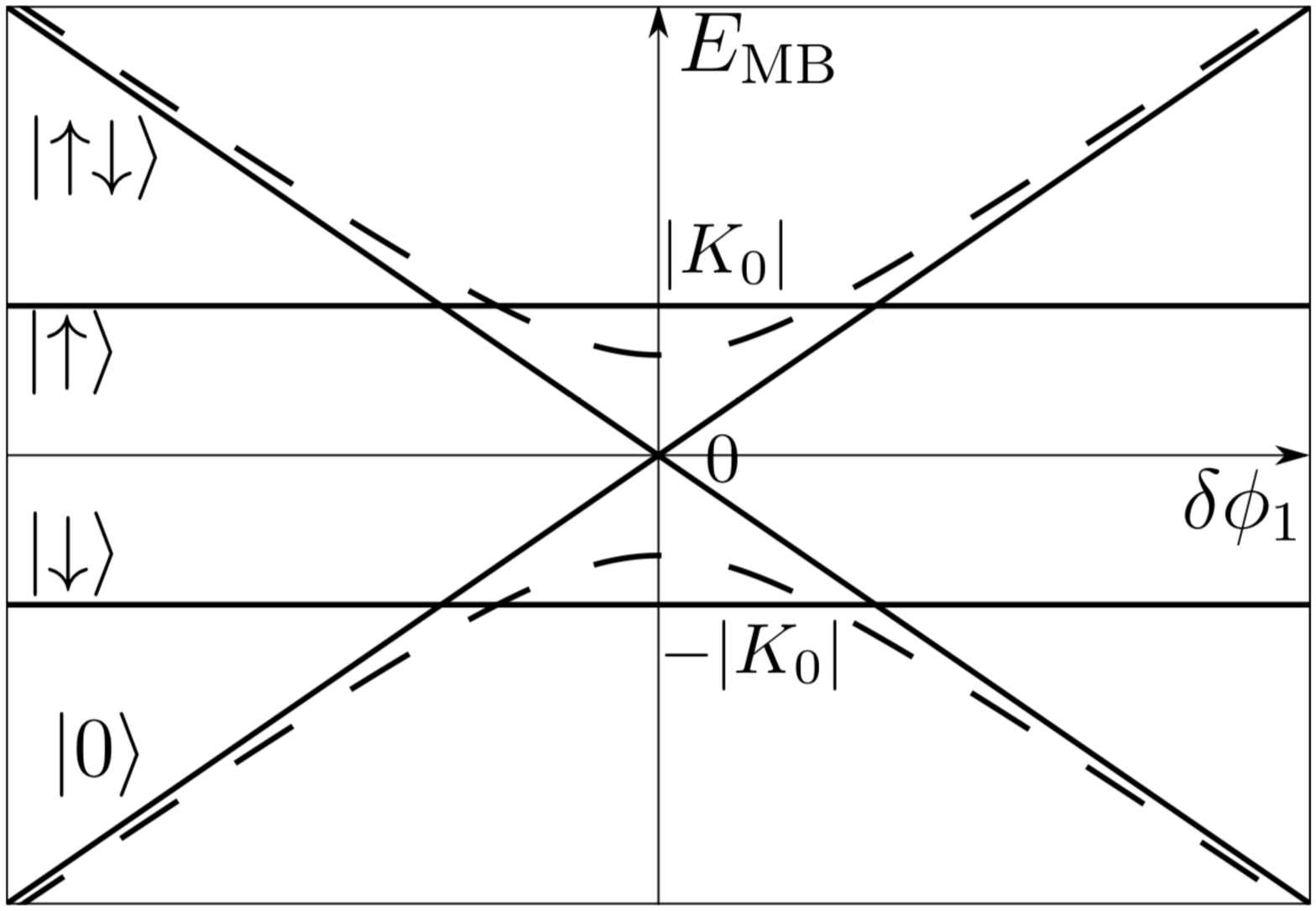}
\caption{Many-body energy spectrum $E_{\mathrm{MB}}$ given by \eqref{MBenergy} corresponding to FIG. \ref{levelSO}.  The ground singlet state, single quasiparticle states of different spin  and the excited singlet are labeled as $\left | 0 \right\rangle$, $\left | \downarrow(\uparrow) \right\rangle$ and $\left | \uparrow\downarrow \right\rangle$, respectively.  The solid (dashed) lines correspond to the ABS plots FIG. \ref{levelSO} a (FIG. \ref{levelSO}b). As the phase is varied, the ground state parity transition between $\left | 0 \right\rangle$ state and $\left | \downarrow \right\rangle$ state takes place at the point defined by \eqref{ellip}.}
\label{fourbody}
\end{center}
\end{figure}

To explain this, and eventually restore the topological protection of $C_{12}$, let us consider many-body states in the vicinity of the Weyl point. 
Their energies are given by the eigenvalues of the many-body Hamiltonian $H_{\rm MB}$ 
\begingroup
\allowdisplaybreaks
\begin{align}
H_{\rm MB}=E_\uparrow(\hat{n}_\uparrow-\frac{1}{2})+E_\downarrow(\hat{n}_\downarrow-\frac{1}{2})
\label{MBenergy}
\end{align}
\endgroup%
where $E_{\uparrow(\downarrow)} = |\vec{h}| \pm |K_0|$ are the energies of quasiparticle excitations with spin up(down), $\hat{n}_{\uparrow(\downarrow)}$ are the number operators of the quasiparticles with the corresponding spin. The energy spectrum $E_{\rm MB}$ for each of the four possible states is given in Fig. \ref{fourbody}. As we see from the Figure, the ground state of the superconducting nanostructure corresponds to $
{n}_{\uparrow}={n}_{\downarrow}=0$ at $|\vec{h}|>|K_0|$ and to 
${n}_{\downarrow}=1, {n}_{\uparrow}=0$ within the ellipsoid $|\vec{h}|<|K_0|$. These states differ in fermion parity, that is the conserving quantity for the superconducting Hamiltonian. This is why the parity transition that takes place at $|\vec{h}|=|K_0|$ is accompanied by the discontinuity of the wave functions, which violates the topological quantization of $C^{12}$. It is evident from 
Fig. \ref{fourbody} that the states of the odd fermion parity do not depend on phases in the vicinity of the Weyl point therefore corresponding to zero $B^{12}$.

The topological protection is restored if one considers the ground state at fixed parity. Then for the even ground state $C^{12}$ is the same as for the spin-degenerate case and experiences an integer jump when the integration plane passes the singularity point.
No change of topological charge occurs for the odd ground state and it remains topologically trivial.

\section{Energy-dependent $S-$matrix\label{Sec:depend}}
In this Section we consider the effect of the energy dependence of the S-matrix on $B_{12}$ given by $\eqref{result1}$ for arbitrary relation between the energy scales of the scattering matrix and the gap $|\Delta|$. 

We make use of the following model scattering matrix:
\begin{equation}
S_\epsilon = \dfrac{i\epsilon - \mu - \mathscr{E}(\hat{H} + i\hat{\Gamma}/2)}{i\epsilon - \mu - \mathscr{E}(\hat{H} - i\hat{\Gamma}/2)},\quad [\hat{H},\hat{\Gamma}]=0
\label{Sdependent}
\end{equation}
where $\hat{\Gamma}, \hat{H}$ are Hermitian dimensionless matrices with eigenvalues of the order of one. This expression can be regarded as a rather general polar decomposition of an energy-dependend scattering matrix. Since the matrices $\hat{\Gamma}, \hat{H}$ can be diagonalized simultaneously, the expession has poles at the complex energies $E=\mu + \mathscr{E}({H}_{n} - i{\Gamma}_n/2)$ defined by the corresponding eigenvalues. The poles can be seen as the scattering resonances. The eigenvalues $H_n$ set the energies of those resonances and the corresponding eigenvalues  $\Gamma_n$ give the inverse lifetimes of these resonances, $\Gamma_n$ must be positive to assure the correct causal properties of the scatterimg. Real energy scale $\mathscr{E}$  then sets the typical spread of the poles in energy around their average position $\mu$.  We note that $S_\epsilon \to 1$ as $\epsilon \to \infty$, so the conditions of regularization described in a previous Section are fulfilled and the integral of $B_{12}$ over a compact subspace in phase space that does not cross the Weyl singularities, reduces to an integer. We remind that the limit $S_\epsilon \to 1$ corresponds to isolated terminals.
\begin{figure}
	\centerline{\includegraphics[width=0.5\textwidth]{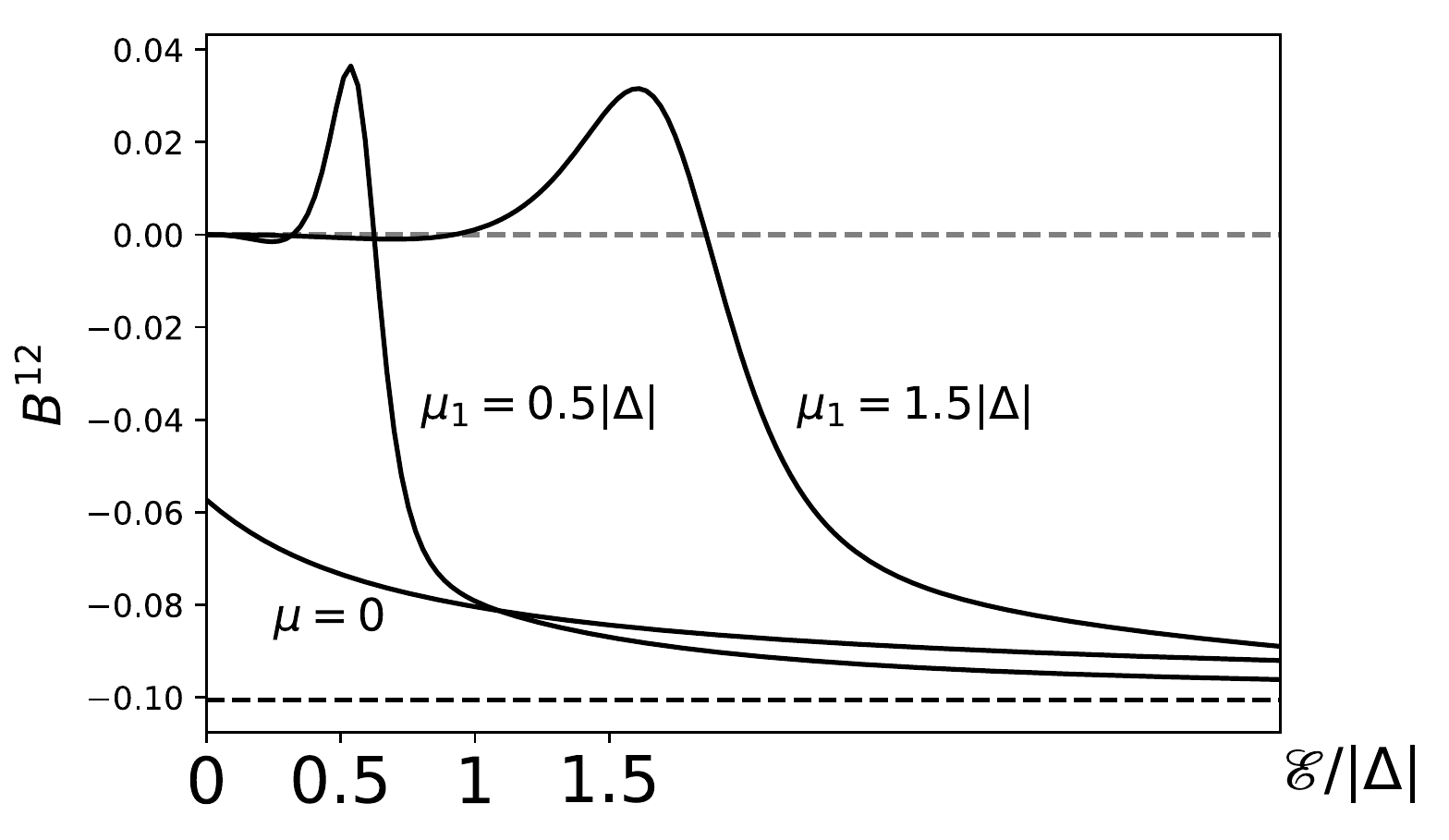}
	}
	\caption{An example plot of $B_12$ (Eq. \ref{result1} for a randomly chosen energy-dependent $S$ versus the energy scale $\mathscr{Epsilon}$ for several choices of the energy scale $\mu$ at $\phi_1 =0.22\pi, \phi_2=-0.67\pi,\phi_3=-\pi$.  The dashed line gives  the limiting value of $B_{12}$ at $\mathscr{Epsilon}\gg|\Delta|$ where the energy dependence of the scattering matrix is weak.   }
	\label{Fig:0not_mu}
\end{figure}
In distinction from the weak energy dependence case, the ABS energies defined by Eq. $\eqref{beenakker}$ can not be readily obtained and the resulting spectrum may be complicated with more ABS per transport channel. It is no more plausible to separate the contributions to $B_{\alpha\beta}$ coming from discrete and  continuous spectrum. This, however, does not change the qualitative features of these contributions discussed above.

Let us consider and illustrate the dependence of $B_{12}$ on these two energy scales. We choose random matrices $\hat{H}$, $\hat{\Gamma}$ that satisfy the conditions stated, and compute $B_{12}$ from Eq. \ref{result1} at rather arbitrary settings of 3 phases. The integration over the imaginary energy in Eq. \ref{result1} permits the evaluation with no regard for the details of a complicated ABS spectrum. We plot the result versus the energy scale $\mathscr{Epsilon}$ at several settings of $\mu$. (Fig. \ref{Fig:0not_mu})

Let us consider $\mu \ne 0$ first. In this case, at $\mathscr{E}\to 0$ the transmission between the terminals is limited to a small circle of the radius $\simeq \mathscr{E}$ near $\mu$. This suppresses the Andreev scattering that requires good transmission at opposite energies, and all quantities that depend on the phase differences including $B_{\alpha\beta}$. In Fig. \ref{Fig:0not_mu}, this is manifested as almost zero $B_{12}$ at $\mathscr{E} < \mu$. The further increase of $\mathscr{E}$ restores the Andreev scattering bringing $B_{12}$ to its typical values of $\sim(2\pi)^{-2}$. We note a non-monotonous dependence on $\mathscr{E}$ and explain it by the fact that different poles of the scattering matrix contribute to $B_{12}$ with typically different signs, and the magnitude of the contribution depends on the position of the pole with respect to the energy scale $\simeq \Delta$.
At $\mathscr{E} \gg \Delta$ the energy dependence of the scattering matrix is weak at $\epsilon \simeq \Delta$ and  $B_{12}$ saturates at a value that does not depend on $\mu$ and is given by Eqs. $\eqref{result2}$ and $\eqref{Large}$ (dashed line in the Figure \ref{Fig:0not_mu}). 

The case of $\mu=0$ is special at small $\mathscr{E}$ since the concentration of transmission in a small circle of energies does not suppress the Andreev scattering. The ABS in this case are concentrated in this small energy circle (see \cite{GolubovReview}) and depend on all phases. This is why $B_{12}$ does not drop to 0 but rather approaches a finite limit at $\mathscr{E} \to 0$.
At $\mathscr{E} \gg \Delta$ $B_{12}$ still saturates at the value corresponding to the weak energy dependence case.

\section{Summary and Conclusions \label{Sec:Sum}}

In this Article, we address the topological properties of multi-terminal superconducting 
nanostructures. This involves Berry curvatures in the parametric space of the superconducting phases of the terminals and associated Chern numbers that manifest themselves in quantized transconductances \cite{ncomms11167}.

The specifics of the superconducting nanostuctures is the presence of continuous spectrum along with the discrete one. The Berry curvature is readily defined for a discrete spectrum. Its generalization for a (partly) continuous spectrum is not straightforward, and is a problem of general interest. It has not been solved in Ref. \cite{ncomms11167}.

We perform the calculation in imaginary time, and model the nanostructure with an energy-dependend scattering matrix. We have derived a general action of superconducting nanostructure with time-dependent phases, this is a separate advance. We expand the action near a point in the space of phases to compute the response function at finite frequency. We define the tensor quantity $B_{\alpha\beta}$ (Eq. \ref{result1}) as a first term in the expansion of the response function at small frequency. This quantity would have been Berry curvature if the spectrum were entirely discrete.

We analyze the topological properties of the computed quantity. Like for Berry curvature, the topological charge associated with divergence of $B_{\alpha\beta}$ is concentrated in the singular points of 3d phase space where ABS cross zero energy --- Weyl points. Unlike Berry curvature, the quantity $B_{\alpha\beta}$ has a non-topological contribution that is constant over the space of phases (Eq. \ref{landauer}). This in general adds a non-quantized part to "Chern" numbers defined as integrals of $B_{\alpha\beta}$ over two superconducting phases, and to the corresponding transconductances. This contribution is determined by the scattering matrix at $\epsilon \to \infty$. It vanishes if the scattering matrix without superconducting phases is time-reversible and if the scattering matrix approaches isolation limit $S_\epsilon=1$ at large energies. For an energy-independent scattering matrix, the non-topological term is associated with the anti-symmetrized part of the conductance matrix of the structure in the normal state.

We consider in detail the case of weak energy dependence of the scattering matrix. We separate the contributions of the discrete and continuous spectrum, find them equally important and derive a compact relation for $B_{\alpha\beta}$ (Eq. \ref{result2}).  

We analyze in detail the Berry curvature in the vicinity of Weyl points. We have found a violation of topological protection of "Chern" number in case of weak spin-orbit coupling. This, however, is rather trivially related to the transition between the ground states of different parity near the Weyl point and associated discontinuity of the wave functions. The topological protection is restored if one considers a ground state of a fixed parity.

We also investigate the properties of $B_{\alpha\beta}$ for the scattering matrices that essentially depend on energy at the energy scale $\simeq \Delta$.
\section{Appendix A: derivation of the action \label{Sec:App1}}
In this Appendix, we derive the effective action for a multi-terminal superconducting junction within the scattering approach. We follow the lines of Ref.\cite{Block}.
\begin{figure}
	\centerline{\includegraphics[width=0.42\textwidth]{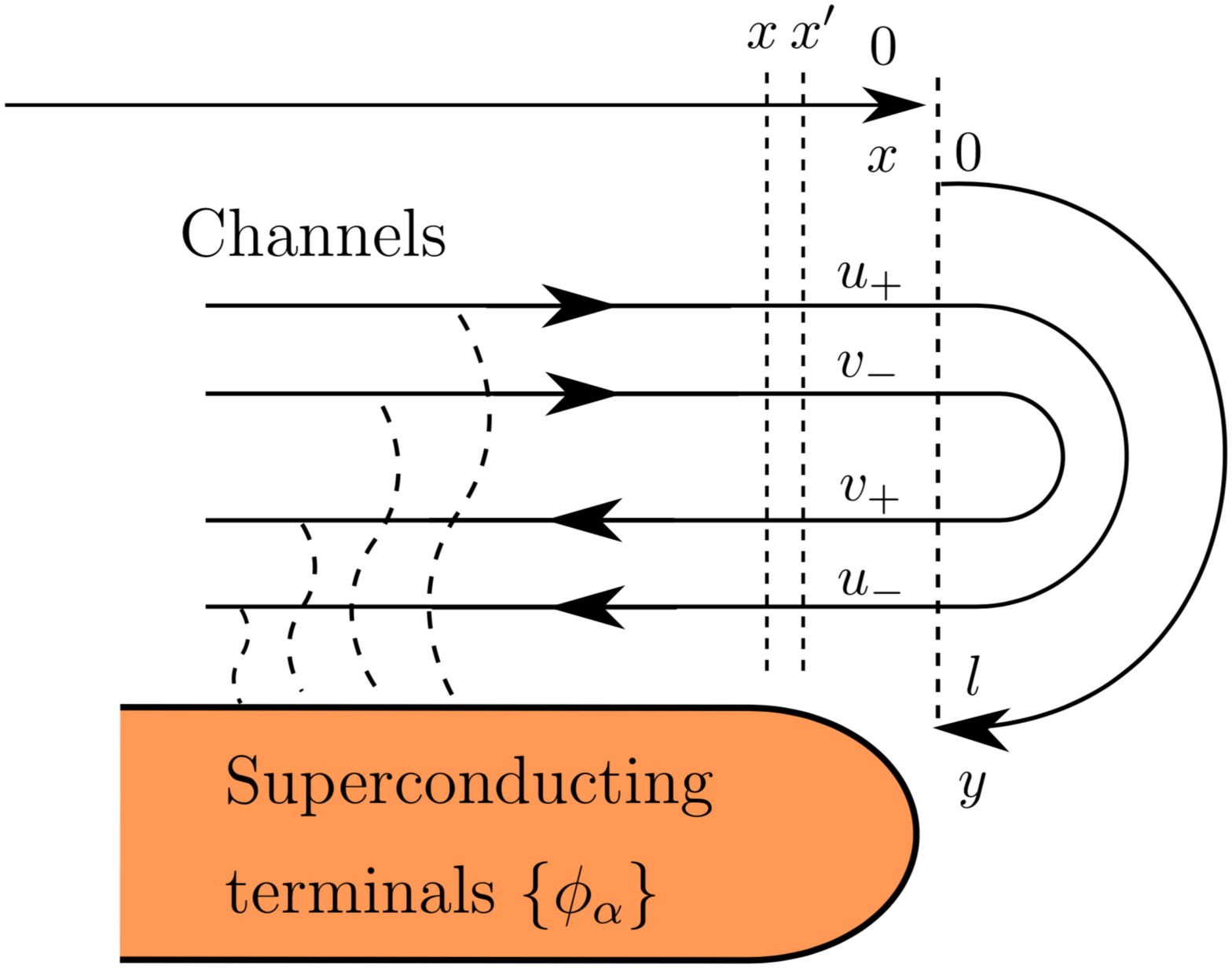}
	}
	\caption{The concrete model for the derivation of the action. The electons are moving in $2N$ spin-degenerate channels connected to the corresponding superconducting terminals by tunneling (wavy dashed lines). In the picture, all the terminals in Eq.$\ref{system}$ are combined into a single superterminal for convenience. Right of the vertical line, the tunnelling between the channels provides the scattering described by $N\times N$ matrix.}
	\label{Fig:Fig1}
\end{figure}
In contrast to Ref.\cite{Block} we proceed in Matsubara formalism. Let us start with the formulation of a concrete microscopic model. Since the scattering formalism is universal, there is a great degree of arbitrariness in the choice of the model: all models that are characterized by the same scattering matrix will result in the same action. Properties of the scatterer are to be completely described by an S-matrix, the details of the model that describes the system are not important. So we choose the model in a way we find it convenient (see Fig. \ref{Fig:Fig1}). We consider a system of independent 1-dimensional channels with pairwise opposite velocities and a linear spectrum. They are defined in the interval $-\infty<x<0$. The total number of channels is $2N$, number $N$ includes the spin doubling. Two channels in a pair with opposite velocities are coupled to the same superconducting reservoir: this is required to assure the time-reversibility of the model at this level. The coupling is a tunnel one, and the coupling strength  is characterized by the dwell time scale $\tau$: at this time scale, an electron in a channel would tunnel to a reservoir. The tunneling results in an addition of self-energy to Green's functions in the channels, which is proportional to the tunneling rate $1/\tau$ and to a matrix Green's function $g$ characterizing a reservoir (see its concrete definition below). The channels defined in such a way model the electron states coming from and going to the reservoirs that are scattered at the nanoscructure. In the scattering region with a coordinate $y\in [0,l]$, there are  $N$ spin-degenerate channels of the same velocity direction. At the boundary $y=0$ the electron amplitudes in the channels match those in the channels of positive velocity at $x=0$ (incoming states), while at $y=l$ the amplitudes match those in the channels with the negative velocity(outgoing states). As we will show, the S-matrix relates the amplitudes at $y=l$ and $y=0$.

To find the action for the nanostructure, we will compute its variation with respect to the variation of $g$. To this end, we require the values of the Green's functions in the channels $x,x'<0$ in close points $x\approx x'$. We find the variation in three steps. At the first step, we express the Green's functions at any $x$ in terms of the Green's functions at $x \approx 0$.  At  the second step, we consider the scattering region that provides a boundary condition. With this, we relate these Green's functions, and solve for them.
This permits to find the variation and the action at the third step.

 In the channels, we choose the basis in the following form 
\begin{equation}
	\begin{pmatrix}
		u_+\\
		v_-\\
		u_-\\
		v_+
	\end{pmatrix}
\end{equation}
where $u_\pm,v_\pm$ are $N$ vectors in the space of the channels associated with the electron and hole amplitudes of the Bogolyubov wave function,  and $\pm$ refers to the sign of the velocity in corresponding channels. In this basis,  the equation for Green's function reads
\begin{equation}
	\left( i\epsilon\tau_3+iv\eta_3\tau_3\partial_x+\frac{i}{2\tau}g\right)G_{\rm Ch}(x,x^\prime)=\delta(x-x^\prime)
	\label{Gequation}
\end{equation}
where $v$ is the velocity that we can set the same for all the channels,
$\epsilon$ is the Matsubara frequency, $\tau_i$ are Pauli matrices in Nambu space, and $\eta_3=\pm$ distinguishes channels with positive and negative velocities. The matrix $g$ is block-diagonal in the channel space. For a given reservoir, it is given by
\begin{equation}
	g=\frac{1}{\sqrt{\epsilon^2+|\Delta|^2}}(\epsilon\tau_3+i\sigma_2[\tau_1(\frac{\Delta-\Delta^*}{2})+i\tau_2(\frac{\Delta+\Delta^\star}{2})]),
	\label{defg}
\end{equation}
 $g^2=1$, $\Delta$ being the superconducting order parameter in the corresponding reservoir. 

We define a block structure
\begin{equation}
	G_{\rm Ch}=\begin{pmatrix}
		G_1 & G_3\\
		G_4 & G_2
	\end{pmatrix}
	\label{blocks}
\end{equation}
We are only interested in the diagonal blocks $G_{1;2}$ since the off-diagonal blocks will not contribute to the variation of the action. We integrate the equation assuming $\epsilon\tau\ll 1$ for $G(x,x')$ at $x<x^\prime$ we obtain
\begin{equation}
	G_{1}(x,x^\prime)=[(\frac{1-g}{2}e^{\frac{(x-x^\prime)}{2v\tau}}+\frac{1+g}{2}e^{-\frac{(x-x^\prime)}{2v\tau}})]G_1^-(x^\prime)
\end{equation}
\begin{equation}
	G_{2}(x,x^\prime)=[(\frac{1+g}{2}e^{\frac{(x-x^\prime)}{2v\tau}}+\frac{1-g}{2}e^{-\frac{(x-x^\prime)}{2v\tau}})]G_2^-(x^\prime)
\end{equation}
where we use special notations for the Green's functions in the close points
\begin{equation}
	G_1^-(x^\prime)=G_1(x^\prime-0,x^\prime),\quad G_2^-(x^\prime)=G_2(x^\prime-0,x^\prime)
\end{equation}
Since the solution for the Green's function should not grow $x\to -\infty$, these Green's functions should satisfy the following conditions
\begin{equation}
	\Pi_+G_{1}^-=0,\quad \Pi_+=\frac{1+g}{2},\quad G_1^-=\lim_{x^\prime\to -0}G_1^-(x^\prime)
	\label{eq1}
\end{equation}
\begin{equation}
	\Pi_-G_2^-=0,\quad \Pi_-=\frac{1-g}{2},\quad G_2^-=\lim_{x^\prime\to -0}G_2^-(x^\prime)
	\label{eq2}
\end{equation}

These matrices $G_{1;2}^-$ can be fixed if we consider the boundary conditions, that can be obtained by solving the equations for the Green's functions in  the the scattering region $y\in [0;l]$. 
To derive these condition, let us introduce the amplitude vectors $\Psi(y) = G(y,x)$, $X(y)=G(x,y)$
that have Nambu structure $\begin{pmatrix}
u(y)\\
v(y)
\end{pmatrix}$ and satisfy the equations
\begin{equation}
	\left(i\epsilon\tau_3+iv\tau_3\partial_y-\begin{pmatrix}
		U(y) & 0\\
		0 & U^T(y)
	\end{pmatrix}\right)\Psi(y)=0
	\label{ampleq}
\end{equation}
\begin{equation}
	\left(i\epsilon\tau_3-iv\tau_3\partial_{y^\prime}\right)X(y^\prime)-X(y^\prime)\begin{pmatrix}
		U(y^\prime) & 0\\
		0 & U^T(y^\prime)
	\end{pmatrix}=0
	\label{ampleq1}
\end{equation}
where $U(y,\epsilon)$ is the $N\times N$ matrix potential acting on electrons inside the scattering region and mixing different channels. The solution of the Eq.$\eqref{ampleq1}$ gives a linear relation on the amplitudes
\begin{equation}
X(y=l)=X(y=0)\hat{S}^\dagger_{-\epsilon}
\label{bncxi}
\end{equation}
where we define the S-matrix for electrons and holes arranged in Nabmu structure 
\begin{equation}
\hat{S}_\epsilon=\begin{pmatrix}
S_e(\epsilon) & 0\\
0 & (S_h(\epsilon))^{-1}
\end{pmatrix}
\end{equation}
\begin{equation}
(S_h(\epsilon))^{-1}\equiv S_{-\epsilon}^T.
\end{equation}
The electron scattering matrix is given by
\begin{equation}
S_e(\epsilon)=S_\epsilon=e^{-\frac{\epsilon l}{v}}\times T_y e^{-\frac{i}{v}\int_{0}^{l}dy U(y,\epsilon)}
\end{equation}
where $T_y$ implies the ordering of the $U(y)$ operators in the exponent according to the values of $y$ in the increasing order. We do not need to specify the energy dependence fo the S-matrix except for the general condition $S_\epsilon S_{-\epsilon}^\dagger=1$.

The relation on the amplitude $\eqref{bncxi}$ gives the relation between the diagonal and off-diagonal blocks of the Green's function $\eqref{blocks}$ outside the scattering region but close to it $|x\epsilon/v|\ll1,|x^\prime\epsilon/v|\ll 1$
\begin{equation}
G_3(x,x)=G_1(x,x^\prime)\hat{S}_{-\epsilon}^\dagger=G_1^- \hat{S}_{-\epsilon}^\dagger,\quad (x<x^\prime)
\label{13}
\end{equation}
The solution of Eq. $\eqref{ampleq}$
\begin{equation}
\Psi(y=l)=\hat{S}_\epsilon\Psi(y=0)
\label{bncpsi}
\end{equation}
yields another relation between the blocks
\begin{equation}
G_2(x^\prime,x)=G_2^+=\hat{S}_\epsilon G_3(x,x),\quad (x<x^\prime)
\label{32}
\end{equation}
 Combining Eq. $\eqref{32}$ and $\eqref{13}$ we obtain the required boundary condition that relates the diagonal sub-blocks
\begin{equation}
	\hat{S}_\epsilon G_1^-\hat{S}_{-\epsilon}^\dagger=G_2^+
	\label{boundc}
\end{equation}
Combining the equations $\eqref{boundc}$, $\eqref{eq1}$ and $\eqref{eq2}$,  and the condition
\begin{equation}
	G^+_{\rm Ch}-G^-_{\rm Ch}=-\frac{i}{v}\tau_3\eta_3
\end{equation}
that  follows directly from $\eqref{Gequation}$ we solve the complete linear system of the equations to obtain the follwing for the diagonal blocks of the general Green's function $\eqref{blocks}$ 
\begin{equation}
	G_1^-=\frac{i}{v}\frac{1}{\Pi_++\Pi_-\hat{S}_\epsilon}\Pi_-\hat{S}_\epsilon,\quad G_1^+=\frac{-i}{v}\frac{1}{\Pi_++\Pi_-\hat{S}_\epsilon}\Pi_+
	\label{G1}
\end{equation}
\begin{equation}
	G_2^-=\frac{-i}{v}\frac{1}{\Pi_-+\Pi_+\hat{S}_{-\epsilon}^\dagger}\Pi_+\hat{S}_{-\epsilon}^\dagger,\quad G_2^+=\frac{i}{v}\frac{1}{\Pi_-+\Pi_+\hat{S}_{-\epsilon}^\dagger}\Pi_-
	\label{G2}
\end{equation}

Next, we employ the formula that expresses the action variation in terms of Green's functions. We vary the reservoir Green's function $g$ keeping normalization $g^2=1$, so that $\{g,\delta g\}=0$, then the variation of the action $L$ is
\begin{equation}
	\delta L=\int dx {\rm Tr}[\delta \Sigma(x) G_{\rm Ch}(x,x)]
\end{equation}
where $\delta \Sigma=\frac{-i}{2\tau}\delta g$ is the variation of self-energy of electrons in channels and $G_{\rm Ch}(x,x)$ is their Green's function at coinciding points. We note here that indeed only the diagonal blocks $G_{1;2}$ in Eq.$\eqref{blocks}$ contribute since $\Sigma$ is diagonal in this basis. The contribution from the channels corresponding to $G_1$ gives
\begin{equation}
	2\delta L_{in}=+\int_{-\infty}^{0}dx{\rm Tr}[\delta \Sigma G_{\rm Ch}(x,x)]=
	\notag
\end{equation}
\begin{equation}
	=\frac{-i}{2\tau}\int_{-\infty}^{0}dx {\rm Tr}[\delta g G_{\rm Ch}(x,x)]=\frac{-1}{2}{\rm Tr}[\delta g \frac{1}{\Pi_++\Pi_- \hat{S}_\epsilon}\Pi_+]
\end{equation}
The futher calculations is convenient to do in the basis that diagonalizes $g$. In this basis, 
\begin{equation}
	\delta g=\begin{pmatrix}
		0 & V \\
		W & 0
	\end{pmatrix},\quad g=\begin{pmatrix}
	1 & 0\\
	0 & -1
\end{pmatrix},\quad \hat{S}=\begin{pmatrix}
S_1 & S_2 \\
S_3 & S_4
\end{pmatrix}
\notag
\end{equation}
\begin{equation}
	\quad Y^{-1}(g+\delta g)Y=g,\quad Y(\hat{S}+\delta \hat{S})Y^{-1}=\hat{S}
\end{equation}
we find
\begin{equation}
	Y=\begin{pmatrix}
		1 & \frac{-V}{2}\\
		\frac{W}{2} & 1
	\end{pmatrix},\quad \delta S_4=-S_3\frac{V}{2}-\frac{W}{2}S_2
	\notag
\end{equation}
\begin{equation}
	2\delta L_{in}=\frac{1}{2}{\rm Tr}VS_4^{-1}S_3
\end{equation}
where all the realtions are valid up to the first order in variations. The contribution from the outgoing channels reads
\begin{equation}
	2\delta L_{out}=\frac{1}{2}{\rm Tr}[\delta g\frac{1}{\Pi_-+\Pi_+\hat{S}_{-\epsilon}^\dagger}\Pi_-]=
	\notag
\end{equation}
\begin{equation}
	=\frac{1}{2}{\rm Tr}[\delta gS_\epsilon\frac{1}{\Pi_-S_\epsilon+\Pi_+}\Pi_-]=\frac{1}{2}{\rm Tr}W S_2 S_4^{-1}
\end{equation}
Summing both contributions, we obtain
\begin{equation}
	2\delta L=-{\rm Tr}[\delta S_4 S_4^{-1}]
\end{equation}
Hence
\begin{equation}
	2L=-{\rm Tr}\log S_4=-{\rm Tr}\log[\Pi_++\Pi_-\hat{S}_\epsilon]
	\label{action}
\end{equation}
This so-called block-determinant result for the action is similar to the one obtained previously \cite{Block} within the Keldysh formalism.

\section{Appendix B: derivation of the response function \label{Sec:App2}}
In this Appendix, we present the details of the derivation of the Eq.$\eqref{general}$ and Eq.$\eqref{result1}$. We start with the action as given by Eq. $\eqref{lagrange}$. In order to derive the response function, we assume that the time-dependent deviation ($\delta \phi(\tau)$) from the stationary phase denoted as $\phi$ is small ($\delta\phi(\tau)\ll 2\pi$) so we can expand the action in Taylor series in $\delta \phi(\tau)$. We also note that in time representation the total phase operator is diagonal ($\phi_{\tau \tau^\prime}=\delta_{\tau \tau^\prime}\phi(\tau)$), which implies that the energy representation of $\phi$ reads
\begin{equation}
	\phi_{nm}=\phi(\omega),\quad\omega=\epsilon_n-\epsilon_m
\end{equation}
We consider here the general case of the energy-dependent scattering matrix. The action from Eq.$\eqref{lagrange}$ reads
\begin{equation}
	-2L={\rm Tr}\log[B+B^T],\quad B=A_\epsilon e^{\frac{-i\phi}{2}}S_\epsilon e^{\frac{i\phi}{2}}A_\epsilon
\end{equation}
$T$ implies the complete operator transposition that includes the reversing of the sign of energy. We remind the definition 
\begin{equation}
A_\epsilon=\sqrt{\frac{E+\epsilon}{2 E}},\quad E=\sqrt{\epsilon^2+|\Delta|^2},
\label{defA}
\end{equation}
We ascribe the stationary part of the phases to an S-matrix $S_\epsilon\to S_\epsilon(\phi)$ and expand in small nonstationary deviation $\delta\phi(\tau)$.
\begin{equation}
	B\simeq B_0+B_1+B_2=B_0+\frac{\partial B}{\partial \phi_\omega^\alpha}\delta\phi_\omega^\alpha+\frac{1}{2}\frac{\partial^2 B}{\partial \phi_\omega^\alpha \partial \phi_{-\omega}^\beta}\delta\phi_\omega^\alpha \delta\phi_{-\omega}^\beta
\end{equation}
We introduce
\begin{equation}
	Q_\epsilon=B_0+B_0^T=A^2_\epsilon S_\epsilon+A^2_{-\epsilon}S^T_{-\epsilon}
\end{equation}
With this,
\begin{align}
	&\delta {\rm Tr}\log[B+B^T]\simeq {\rm Tr}Q^{-1}(B_1+B_1^T+B_2+B_2^T)-\notag\\
	&\frac{1}{2}{\rm Tr}Q^{-1}(B_1+B_1^T)Q^{-1}(B_1+B_1^T).
	\label{expansion}
\end{align}
We remind the definition of the matrix, that  projects on the channels connected to a given terminal $\alpha$:
\begin{equation}
	(I^\alpha)^{ab}=\delta^{ab}\begin{cases}
		1,\quad a=\alpha \\
		0,\quad a\ne \alpha
	\end{cases}
\end{equation}
where $a,b$ indices are in channels. With the help of this matrix the phase variation can be conveniently expressed as
\begin{equation}
	(\delta \phi^\alpha)^{ab}= (I^\alpha)^{ab}\delta\phi^\alpha(\tau)
\end{equation}
 For simplicity of the notations, we denote the stationary phase derivatives $\partial_{\phi_\alpha}=\partial_\alpha$. With all this we consider the expansion of the S-matrix
\begin{align}
	&e^{\frac{-i\delta\phi(\tau)}{2}}S_\epsilon e^{\frac{i\delta\phi(\tau)}{2}}\simeq S_\epsilon+[S_\epsilon,\frac{i\delta\phi(\tau)}{2}]+\frac{\delta\phi(\tau)}{2}S_\epsilon \frac{\delta\phi(\tau)}{2}-\notag\\
	&-\frac{1}{2}\{(\frac{\delta\phi(\tau)}{2})^2,S_\epsilon\}
\end{align}
Let us we also note the identities for the derivatives with respect to the stationary phases:
\begin{equation}
	\frac{\partial S}{\partial \alpha}=[S,\frac{iI_\alpha}{2}]
\end{equation}
\begin{equation}
	\frac{\partial^2 S}{\partial \alpha\partial \beta}=\frac{I_\alpha}{2}S\frac{I_\beta}{2}+\frac{I_\beta}{2}S\frac{I_\alpha}{2}-\delta_{\alpha \beta}\{\frac{I_\alpha}{4},S\}
\end{equation}
the first term in the expansion $\eqref{expansion}$ vanishes since $\delta\phi_{\omega=0}=0$. The second term is
\begin{align}
	&{\rm Tr}Q^{-1}(B_2+B_2^T)=2{\rm Tr}Q^{-1}B_2=\notag\\
	&\delta\phi_\omega^\alpha \delta\phi_{-\omega}^\beta \int \frac{d\epsilon}{2\pi} {\rm Tr}Q_\epsilon^{-1}A_{\epsilon}^2[-\delta_{\alpha \beta}\{\frac{I_\alpha}{4},S_{\epsilon}\}+\notag\\
	&+\frac{I_\alpha}{2}S_{\epsilon-\omega}\frac{I_\beta}{2}+\frac{I_\beta}{2}S_{\epsilon+\omega}\frac{I_\alpha}{2}]=
	\notag\\
	&=\frac{\delta\phi^\alpha_\omega \delta\phi^\alpha_{-\omega}}{2}\int \frac{d\epsilon}{2\pi} {\rm Tr}Q_\epsilon^{-1}[\frac{\partial^2 Q_\epsilon}{\partial \alpha \partial \beta}]+\notag\\
&\delta\phi^\alpha_\omega \delta\phi^\alpha_{-\omega}\int \frac{d\epsilon}{2\pi} {\rm Tr} Q_\epsilon^{-1}A_\epsilon^2[\frac{I_\alpha}{2}(S_{\epsilon-\omega}-S_\epsilon)\frac{I_\beta}{2}+\notag\\
	&\frac{I_\beta}{2}(S_{\epsilon+\omega}-S_\epsilon)\frac{I_\alpha}{2}]
	\label{part1R}
\end{align}
The first term here does not depend on frequency and does not vanish in the limit $\omega\to 0$. The second term up to linear order in $\omega$ can be rewritten as
\begin{align}
	&2\omega\delta\phi^\alpha_\omega \delta\phi^\alpha_{-\omega}\int \frac{d\epsilon}{2\pi} {\rm Tr}[Q_\epsilon^{-1}A_\epsilon^2 \frac{I_\beta}{2}\frac{\partial S_\epsilon}{\partial \epsilon}\frac{I_\alpha}{2}]=\notag\\
	&\omega \delta\phi^\alpha_\omega \delta\phi^\alpha_{-\omega}\int \frac{d\epsilon}{2\pi} {\rm Tr} Q_\epsilon ^{-1} A_\epsilon^2 \partial_\beta \{\frac{\partial S_\epsilon}{\partial \epsilon},\frac{iI_\alpha}{2}\}
	\label{first}
\end{align}
The second term in the expansion $\eqref{expansion}$ reads
\begin{align}
	&-\frac{1}{2}{\rm Tr}Q^{-1}(B_1+B_1^T)Q^{-1}(B_1+B_1^T)=\notag\\
	&\frac{-\delta\phi^\alpha_\omega \delta\phi^\beta_{-\omega}}{2}\int \frac{d\epsilon}{2\pi} {\rm Tr} Q^{-1}_1(A_{-1}(\frac{i I_\alpha}{2}S_{-2}^T-S_{-1}^T\frac{i I_\alpha}{2})A_{-2}-\notag \\
	&A_{1}(\frac{i I_\alpha}{2}S_{2}-S_{1}\frac{i I_\alpha}{2})A_{2}) Q^{-1}_2(A_{-2}(\frac{i I_\beta}{2}S_{-1}^T-\notag\\
	&S_{-2}^T\frac{i I_\beta}{2})A_{-1}-A_{2}(\frac{i I_\beta}{2}S_{1}-S_{2}\frac{i I_\beta}{2})A_{1})
\end{align}
where subscripts mean taking the function at the frequency $\epsilon_{1,2}:\quad \epsilon_1=\epsilon_2+\omega$ and we denoted $\epsilon_2=\epsilon$. Summing it with $\eqref{part1R}$ we get the general response function as in Eq. $\eqref{general}$.

To perform the adiabatic expansion in the small parameter $\omega/|\Delta|$ here we keep $\omega$ as an independent parameter. We will use the identities
\begin{equation}
	\frac{iI_\alpha}{2}S_2-S_1\frac{iI_\alpha}{2}=-\frac{\partial S_{cl}}{\partial \alpha}-\{S_q,\frac{iI_\alpha}{2}\}
\end{equation}
where we introduced "classical" and "quantum" S-matrices as
\begin{equation}
	S_{cl}=\frac{S_1+S_2}{2},\quad S_q=\frac{S_1-S_2}{2}
\end{equation}
With this, we rewrite the term
\begin{align}
	&\frac{-\delta\phi^\alpha_\omega \delta\phi^\beta_{-\omega}}{2}\int \frac{d\epsilon}{2\pi} {\rm Tr} Q_1^{-1}[A_1 A_2(\frac{\partial S_{cl}}{\partial \alpha}+\{S_q,\frac{iI_\alpha}{2}\})+\notag \\
	&A_{-1} A_{-2}(\frac{\partial S^T_{cl}}{\partial \alpha}-\{S_q,\frac{iI_\alpha}{2}\})]Q_2^{-1}[A_1 A_2(\frac{\partial S_{cl}}{\partial \alpha}-\{S_q,\frac{iI_\alpha}{2}\})\notag\\
	&+A_{-1} A_{-2}(\frac{\partial S^T_{cl}}{\partial \alpha}+\{S_q,\frac{iI_\alpha}{2}\})]
\end{align}
Next, we expand the terms that are taken at $\epsilon_1=\epsilon_2+\omega$. Thery come from three factors here. The expansion of the first factor $Q^{-1}_1\simeq Q_2^{-1}+\omega \frac{\partial Q_\epsilon^{-1}}{\partial \epsilon}$ gives rise to
\begin{equation}
	\frac{\omega}{2}\delta\phi^\alpha_\omega \delta\phi^\beta_{-\omega}\int \frac{d\epsilon}{2\pi}{\rm Tr}Q^{-1}_\epsilon\frac{\partial Q_\epsilon}{\partial \epsilon}Q^{-1}_\epsilon\frac{\partial Q_\epsilon}{\partial \alpha}Q^{-1}_\epsilon\frac{\partial Q_\epsilon}{\partial \beta}
\end{equation}
The expansion of the product of the classical parts is symmetric with respect to $\alpha,\beta$, so it vanishes. The product of quantum parts vanishes in linear order in $\omega$. So we only need to consider quantum times classical and expand the quantum one
\begin{equation}
	S_q\simeq \frac{\omega}{2} \frac{\partial S_\epsilon}{\partial \epsilon}
\end{equation}
it yields
\begin{align}
	&-\frac{2}{2}\delta\phi^\alpha_\omega \delta\phi^\beta_{-\omega}\int \frac{d\epsilon}{2\pi} {\rm Tr}Q^{-1}\frac{\omega}{2}(A_\epsilon^2\{\frac{iI_\alpha}{2},\frac{\partial S_\epsilon}{\partial \epsilon}\}-\notag\\
	&-A_{-\epsilon}^2\{\frac{iI_\alpha}{2},\frac{\partial S^T_{-\epsilon}}{\partial \epsilon}\})\frac{\partial Q}{\partial \beta}Q^{-1}=\notag\\
	&=\omega \delta\phi^\alpha_\omega \delta\phi^\beta_{-\omega}\int \frac{d\epsilon}{2\pi} {\rm Tr} \frac{\partial Q^{-1}}{\partial \beta}A_\epsilon^2\{\frac{\partial S_\epsilon}{\partial \epsilon},\frac{iI_\alpha}{2}\}
\end{align}
where the first doubling is due to the same contribution with $\alpha \leftrightarrow \beta$.
Summing it with $\eqref{first}$ we obtain the total response function as given by $\eqref{result1}$
\begin{align}
	&-\frac{2_S}{2}\omega\delta\phi^\alpha_\omega \delta\phi^\beta_{-\omega}\int \frac{d\epsilon}{2\pi}(\frac{1}{2}{\rm Tr}Q^{-1}_\epsilon\frac{\partial Q_\epsilon}{\partial \epsilon}Q^{-1}_\epsilon\frac{\partial Q_\epsilon}{\partial \alpha}Q^{-1}_\epsilon\frac{\partial Q_\epsilon}{\partial \beta}+\notag\\
	&\frac{\partial}{\partial \beta}{\rm Tr}[Q_\epsilon^{-1}A^2(\epsilon)\{\frac{\partial S_\epsilon}{\partial \epsilon},\frac{i I_\alpha}{2}\}])
\end{align}

\begin{acknowledgements}
This project has received funding from the European Research Council (ERC) under the European Union's Horizon 2020 research and innovation programme (grant agreement \textnumero 694272).
\end{acknowledgements}

\bibliographystyle{apsrev4-1}
\bibliography{ManusciptRepinChenNazarov}

\begin{thebibliography}{48}%
\makeatletter
\providecommand \@ifxundefined [1]{%
 \@ifx{#1\undefined}
}%
\providecommand \@ifnum [1]{%
 \ifnum #1\expandafter \@firstoftwo
 \else \expandafter \@secondoftwo
 \fi
}%
\providecommand \@ifx [1]{%
 \ifx #1\expandafter \@firstoftwo
 \else \expandafter \@secondoftwo
 \fi
}%
\providecommand \natexlab [1]{#1}%
\providecommand \enquote  [1]{``#1''}%
\providecommand \bibnamefont  [1]{#1}%
\providecommand \bibfnamefont [1]{#1}%
\providecommand \citenamefont [1]{#1}%
\providecommand \href@noop [0]{\@secondoftwo}%
\providecommand \href [0]{\begingroup \@sanitize@url \@href}%
\providecommand \@href[1]{\@@startlink{#1}\@@href}%
\providecommand \@@href[1]{\endgroup#1\@@endlink}%
\providecommand \@sanitize@url [0]{\catcode `\\12\catcode `\$12\catcode
  `\&12\catcode `\#12\catcode `\^12\catcode `\_12\catcode `\%12\relax}%
\providecommand \@@startlink[1]{}%
\providecommand \@@endlink[0]{}%
\providecommand \url  [0]{\begingroup\@sanitize@url \@url }%
\providecommand \@url [1]{\endgroup\@href {#1}{\urlprefix }}%
\providecommand \urlprefix  [0]{URL }%
\providecommand \Eprint [0]{\href }%
\providecommand \doibase [0]{http://dx.doi.org/}%
\providecommand \selectlanguage [0]{\@gobble}%
\providecommand \bibinfo  [0]{\@secondoftwo}%
\providecommand \bibfield  [0]{\@secondoftwo}%
\providecommand \translation [1]{[#1]}%
\providecommand \BibitemOpen [0]{}%
\providecommand \bibitemStop [0]{}%
\providecommand \bibitemNoStop [0]{.\EOS\space}%
\providecommand \EOS [0]{\spacefactor3000\relax}%
\providecommand \BibitemShut  [1]{\csname bibitem#1\endcsname}%
\let\auto@bib@innerbib\@empty
\bibitem [{\citenamefont {Xu}\ and\ \citenamefont
  {Balents}(2018)}]{PhysRevLett.121.087001}%
  \BibitemOpen
  \bibfield  {author} {\bibinfo {author} {\bibfnamefont {C.}~\bibnamefont
  {Xu}}\ and\ \bibinfo {author} {\bibfnamefont {L.}~\bibnamefont {Balents}},\
  }\href {\doibase 10.1103/PhysRevLett.121.087001} {\bibfield  {journal}
  {\bibinfo  {journal} {Phys. Rev. Lett.}\ }\textbf {\bibinfo {volume} {121}},\
  \bibinfo {pages} {087001} (\bibinfo {year} {2018})}\BibitemShut {NoStop}%
\bibitem [{\citenamefont {Yao}\ and\ \citenamefont
  {Wang}(2018)}]{PhysRevLett.121.086803}%
  \BibitemOpen
  \bibfield  {author} {\bibinfo {author} {\bibfnamefont {S.}~\bibnamefont
  {Yao}}\ and\ \bibinfo {author} {\bibfnamefont {Z.}~\bibnamefont {Wang}},\
  }\href {\doibase 10.1103/PhysRevLett.121.086803} {\bibfield  {journal}
  {\bibinfo  {journal} {Phys. Rev. Lett.}\ }\textbf {\bibinfo {volume} {121}},\
  \bibinfo {pages} {086803} (\bibinfo {year} {2018})}\BibitemShut {NoStop}%
\bibitem [{\citenamefont {Pacholski}\ \emph {et~al.}(2018)\citenamefont
  {Pacholski}, \citenamefont {Beenakker},\ and\ \citenamefont
  {Adagideli}}]{PhysRevLett.121.037701}%
  \BibitemOpen
  \bibfield  {author} {\bibinfo {author} {\bibfnamefont {M.~J.}\ \bibnamefont
  {Pacholski}}, \bibinfo {author} {\bibfnamefont {C.~W.~J.}\ \bibnamefont
  {Beenakker}}, \ and\ \bibinfo {author} {\bibfnamefont {i.~d.~I.}\
  \bibnamefont {Adagideli}},\ }\href {\doibase 10.1103/PhysRevLett.121.037701}
  {\bibfield  {journal} {\bibinfo  {journal} {Phys. Rev. Lett.}\ }\textbf
  {\bibinfo {volume} {121}},\ \bibinfo {pages} {037701} (\bibinfo {year}
  {2018})}\BibitemShut {NoStop}%
\bibitem [{\citenamefont {Hossain}\ \emph {et~al.}(2018)\citenamefont
  {Hossain}, \citenamefont {Ma}, \citenamefont {Mueed}, \citenamefont
  {Pfeiffer}, \citenamefont {West}, \citenamefont {Baldwin},\ and\
  \citenamefont {Shayegan}}]{PhysRevLett.120.256601}%
  \BibitemOpen
  \bibfield  {author} {\bibinfo {author} {\bibfnamefont {M.~S.}\ \bibnamefont
  {Hossain}}, \bibinfo {author} {\bibfnamefont {M.~K.}\ \bibnamefont {Ma}},
  \bibinfo {author} {\bibfnamefont {M.~A.}\ \bibnamefont {Mueed}}, \bibinfo
  {author} {\bibfnamefont {L.~N.}\ \bibnamefont {Pfeiffer}}, \bibinfo {author}
  {\bibfnamefont {K.~W.}\ \bibnamefont {West}}, \bibinfo {author}
  {\bibfnamefont {K.~W.}\ \bibnamefont {Baldwin}}, \ and\ \bibinfo {author}
  {\bibfnamefont {M.}~\bibnamefont {Shayegan}},\ }\href {\doibase
  10.1103/PhysRevLett.120.256601} {\bibfield  {journal} {\bibinfo  {journal}
  {Phys. Rev. Lett.}\ }\textbf {\bibinfo {volume} {120}},\ \bibinfo {pages}
  {256601} (\bibinfo {year} {2018})}\BibitemShut {NoStop}%
\bibitem [{\citenamefont {Tan}\ \emph {et~al.}(2018)\citenamefont {Tan},
  \citenamefont {Zhang}, \citenamefont {Liu}, \citenamefont {Xue},
  \citenamefont {Yu}, \citenamefont {Zhu}, \citenamefont {Yan}, \citenamefont
  {Zhu},\ and\ \citenamefont {Yu}}]{PhysRevLett.120.130503}%
  \BibitemOpen
  \bibfield  {author} {\bibinfo {author} {\bibfnamefont {X.}~\bibnamefont
  {Tan}}, \bibinfo {author} {\bibfnamefont {D.-W.}\ \bibnamefont {Zhang}},
  \bibinfo {author} {\bibfnamefont {Q.}~\bibnamefont {Liu}}, \bibinfo {author}
  {\bibfnamefont {G.}~\bibnamefont {Xue}}, \bibinfo {author} {\bibfnamefont
  {H.-F.}\ \bibnamefont {Yu}}, \bibinfo {author} {\bibfnamefont {Y.-Q.}\
  \bibnamefont {Zhu}}, \bibinfo {author} {\bibfnamefont {H.}~\bibnamefont
  {Yan}}, \bibinfo {author} {\bibfnamefont {S.-L.}\ \bibnamefont {Zhu}}, \ and\
  \bibinfo {author} {\bibfnamefont {Y.}~\bibnamefont {Yu}},\ }\href {\doibase
  10.1103/PhysRevLett.120.130503} {\bibfield  {journal} {\bibinfo  {journal}
  {Phys. Rev. Lett.}\ }\textbf {\bibinfo {volume} {120}},\ \bibinfo {pages}
  {130503} (\bibinfo {year} {2018})}\BibitemShut {NoStop}%
\bibitem [{\citenamefont {Brems}\ \emph {et~al.}(2018)\citenamefont {Brems},
  \citenamefont {Paaske}, \citenamefont {Lunde},\ and\ \citenamefont
  {Willatzen}}]{PhysRevB.97.081402}%
  \BibitemOpen
  \bibfield  {author} {\bibinfo {author} {\bibfnamefont {M.~R.}\ \bibnamefont
  {Brems}}, \bibinfo {author} {\bibfnamefont {J.}~\bibnamefont {Paaske}},
  \bibinfo {author} {\bibfnamefont {A.~M.}\ \bibnamefont {Lunde}}, \ and\
  \bibinfo {author} {\bibfnamefont {M.}~\bibnamefont {Willatzen}},\ }\href
  {\doibase 10.1103/PhysRevB.97.081402} {\bibfield  {journal} {\bibinfo
  {journal} {Phys. Rev. B}\ }\textbf {\bibinfo {volume} {97}},\ \bibinfo
  {pages} {081402} (\bibinfo {year} {2018})}\BibitemShut {NoStop}%
\bibitem [{\citenamefont {Tang}\ \emph {et~al.}(2017)\citenamefont {Tang},
  \citenamefont {Ikushima}, \citenamefont {Ling}, \citenamefont {Chi},\ and\
  \citenamefont {Chen}}]{PhysRevApplied.8.064001}%
  \BibitemOpen
  \bibfield  {author} {\bibinfo {author} {\bibfnamefont {C.-C.}\ \bibnamefont
  {Tang}}, \bibinfo {author} {\bibfnamefont {K.}~\bibnamefont {Ikushima}},
  \bibinfo {author} {\bibfnamefont {D.~C.}\ \bibnamefont {Ling}}, \bibinfo
  {author} {\bibfnamefont {C.~C.}\ \bibnamefont {Chi}}, \ and\ \bibinfo
  {author} {\bibfnamefont {J.-C.}\ \bibnamefont {Chen}},\ }\href {\doibase
  10.1103/PhysRevApplied.8.064001} {\bibfield  {journal} {\bibinfo  {journal}
  {Phys. Rev. Applied}\ }\textbf {\bibinfo {volume} {8}},\ \bibinfo {pages}
  {064001} (\bibinfo {year} {2017})}\BibitemShut {NoStop}%
\bibitem [{\citenamefont {G\"otte}\ \emph {et~al.}(2014)\citenamefont
  {G\"otte}, \citenamefont {Paananen}, \citenamefont {Reiss},\ and\
  \citenamefont {Dahm}}]{PhysRevApplied.2.054010}%
  \BibitemOpen
  \bibfield  {author} {\bibinfo {author} {\bibfnamefont {M.}~\bibnamefont
  {G\"otte}}, \bibinfo {author} {\bibfnamefont {T.}~\bibnamefont {Paananen}},
  \bibinfo {author} {\bibfnamefont {G.}~\bibnamefont {Reiss}}, \ and\ \bibinfo
  {author} {\bibfnamefont {T.}~\bibnamefont {Dahm}},\ }\href {\doibase
  10.1103/PhysRevApplied.2.054010} {\bibfield  {journal} {\bibinfo  {journal}
  {Phys. Rev. Applied}\ }\textbf {\bibinfo {volume} {2}},\ \bibinfo {pages}
  {054010} (\bibinfo {year} {2014})}\BibitemShut {NoStop}%
\bibitem [{\citenamefont {Maciejko}\ \emph {et~al.}(2010)\citenamefont
  {Maciejko}, \citenamefont {Kim},\ and\ \citenamefont
  {Qi}}]{PhysRevB.82.195409}%
  \BibitemOpen
  \bibfield  {author} {\bibinfo {author} {\bibfnamefont {J.}~\bibnamefont
  {Maciejko}}, \bibinfo {author} {\bibfnamefont {E.-A.}\ \bibnamefont {Kim}}, \
  and\ \bibinfo {author} {\bibfnamefont {X.-L.}\ \bibnamefont {Qi}},\ }\href
  {\doibase 10.1103/PhysRevB.82.195409} {\bibfield  {journal} {\bibinfo
  {journal} {Phys. Rev. B}\ }\textbf {\bibinfo {volume} {82}},\ \bibinfo
  {pages} {195409} (\bibinfo {year} {2010})}\BibitemShut {NoStop}%
\bibitem [{\citenamefont {Chen}\ \emph {et~al.}(2011)\citenamefont {Chen},
  \citenamefont {Zhu}, \citenamefont {Xiao},\ and\ \citenamefont
  {Zhang}}]{PhysRevLett.107.056804}%
  \BibitemOpen
  \bibfield  {author} {\bibinfo {author} {\bibfnamefont {H.}~\bibnamefont
  {Chen}}, \bibinfo {author} {\bibfnamefont {W.}~\bibnamefont {Zhu}}, \bibinfo
  {author} {\bibfnamefont {D.}~\bibnamefont {Xiao}}, \ and\ \bibinfo {author}
  {\bibfnamefont {Z.}~\bibnamefont {Zhang}},\ }\href {\doibase
  10.1103/PhysRevLett.107.056804} {\bibfield  {journal} {\bibinfo  {journal}
  {Phys. Rev. Lett.}\ }\textbf {\bibinfo {volume} {107}},\ \bibinfo {pages}
  {056804} (\bibinfo {year} {2011})}\BibitemShut {NoStop}%
\bibitem [{\citenamefont {Nayak}\ \emph {et~al.}(2008)\citenamefont {Nayak},
  \citenamefont {Simon}, \citenamefont {Stern}, \citenamefont {Freedman},\ and\
  \citenamefont {Das~Sarma}}]{RevModPhys.80.1083}%
  \BibitemOpen
  \bibfield  {author} {\bibinfo {author} {\bibfnamefont {C.}~\bibnamefont
  {Nayak}}, \bibinfo {author} {\bibfnamefont {S.~H.}\ \bibnamefont {Simon}},
  \bibinfo {author} {\bibfnamefont {A.}~\bibnamefont {Stern}}, \bibinfo
  {author} {\bibfnamefont {M.}~\bibnamefont {Freedman}}, \ and\ \bibinfo
  {author} {\bibfnamefont {S.}~\bibnamefont {Das~Sarma}},\ }\href {\doibase
  10.1103/RevModPhys.80.1083} {\bibfield  {journal} {\bibinfo  {journal} {Rev.
  Mod. Phys.}\ }\textbf {\bibinfo {volume} {80}},\ \bibinfo {pages} {1083}
  (\bibinfo {year} {2008})}\BibitemShut {NoStop}%
\bibitem [{\citenamefont {Aasen}\ \emph {et~al.}(2016)\citenamefont {Aasen},
  \citenamefont {Hell}, \citenamefont {Mishmash}, \citenamefont {Higginbotham},
  \citenamefont {Danon}, \citenamefont {Leijnse}, \citenamefont {Jespersen},
  \citenamefont {Folk}, \citenamefont {Marcus}, \citenamefont {Flensberg},\
  and\ \citenamefont {Alicea}}]{PhysRevX.6.031016}%
  \BibitemOpen
  \bibfield  {author} {\bibinfo {author} {\bibfnamefont {D.}~\bibnamefont
  {Aasen}}, \bibinfo {author} {\bibfnamefont {M.}~\bibnamefont {Hell}},
  \bibinfo {author} {\bibfnamefont {R.~V.}\ \bibnamefont {Mishmash}}, \bibinfo
  {author} {\bibfnamefont {A.}~\bibnamefont {Higginbotham}}, \bibinfo {author}
  {\bibfnamefont {J.}~\bibnamefont {Danon}}, \bibinfo {author} {\bibfnamefont
  {M.}~\bibnamefont {Leijnse}}, \bibinfo {author} {\bibfnamefont {T.~S.}\
  \bibnamefont {Jespersen}}, \bibinfo {author} {\bibfnamefont {J.~A.}\
  \bibnamefont {Folk}}, \bibinfo {author} {\bibfnamefont {C.~M.}\ \bibnamefont
  {Marcus}}, \bibinfo {author} {\bibfnamefont {K.}~\bibnamefont {Flensberg}}, \
  and\ \bibinfo {author} {\bibfnamefont {J.}~\bibnamefont {Alicea}},\ }\href
  {\doibase 10.1103/PhysRevX.6.031016} {\bibfield  {journal} {\bibinfo
  {journal} {Phys. Rev. X}\ }\textbf {\bibinfo {volume} {6}},\ \bibinfo {pages}
  {031016} (\bibinfo {year} {2016})}\BibitemShut {NoStop}%
\bibitem [{\citenamefont {Kane}\ and\ \citenamefont
  {Mele}(2005)}]{PhysRevLett.95.226801}%
  \BibitemOpen
  \bibfield  {author} {\bibinfo {author} {\bibfnamefont {C.~L.}\ \bibnamefont
  {Kane}}\ and\ \bibinfo {author} {\bibfnamefont {E.~J.}\ \bibnamefont
  {Mele}},\ }\href {\doibase 10.1103/PhysRevLett.95.226801} {\bibfield
  {journal} {\bibinfo  {journal} {Phys. Rev. Lett.}\ }\textbf {\bibinfo
  {volume} {95}},\ \bibinfo {pages} {226801} (\bibinfo {year}
  {2005})}\BibitemShut {NoStop}%
\bibitem [{\citenamefont {Wu}\ \emph {et~al.}(2006)\citenamefont {Wu},
  \citenamefont {Bernevig},\ and\ \citenamefont
  {Zhang}}]{PhysRevLett.96.106401}%
  \BibitemOpen
  \bibfield  {author} {\bibinfo {author} {\bibfnamefont {C.}~\bibnamefont
  {Wu}}, \bibinfo {author} {\bibfnamefont {B.~A.}\ \bibnamefont {Bernevig}}, \
  and\ \bibinfo {author} {\bibfnamefont {S.-C.}\ \bibnamefont {Zhang}},\ }\href
  {\doibase 10.1103/PhysRevLett.96.106401} {\bibfield  {journal} {\bibinfo
  {journal} {Phys. Rev. Lett.}\ }\textbf {\bibinfo {volume} {96}},\ \bibinfo
  {pages} {106401} (\bibinfo {year} {2006})}\BibitemShut {NoStop}%
\bibitem [{\citenamefont {Fu}\ \emph {et~al.}(2007)\citenamefont {Fu},
  \citenamefont {Kane},\ and\ \citenamefont {Mele}}]{PhysRevLett.98.106803}%
  \BibitemOpen
  \bibfield  {author} {\bibinfo {author} {\bibfnamefont {L.}~\bibnamefont
  {Fu}}, \bibinfo {author} {\bibfnamefont {C.~L.}\ \bibnamefont {Kane}}, \ and\
  \bibinfo {author} {\bibfnamefont {E.~J.}\ \bibnamefont {Mele}},\ }\href
  {\doibase 10.1103/PhysRevLett.98.106803} {\bibfield  {journal} {\bibinfo
  {journal} {Phys. Rev. Lett.}\ }\textbf {\bibinfo {volume} {98}},\ \bibinfo
  {pages} {106803} (\bibinfo {year} {2007})}\BibitemShut {NoStop}%
\bibitem [{\citenamefont {Qi}\ \emph {et~al.}(2009)\citenamefont {Qi},
  \citenamefont {Hughes}, \citenamefont {Raghu},\ and\ \citenamefont
  {Zhang}}]{PhysRevLett.102.187001}%
  \BibitemOpen
  \bibfield  {author} {\bibinfo {author} {\bibfnamefont {X.-L.}\ \bibnamefont
  {Qi}}, \bibinfo {author} {\bibfnamefont {T.~L.}\ \bibnamefont {Hughes}},
  \bibinfo {author} {\bibfnamefont {S.}~\bibnamefont {Raghu}}, \ and\ \bibinfo
  {author} {\bibfnamefont {S.-C.}\ \bibnamefont {Zhang}},\ }\href {\doibase
  10.1103/PhysRevLett.102.187001} {\bibfield  {journal} {\bibinfo  {journal}
  {Phys. Rev. Lett.}\ }\textbf {\bibinfo {volume} {102}},\ \bibinfo {pages}
  {187001} (\bibinfo {year} {2009})}\BibitemShut {NoStop}%
\bibitem [{\citenamefont {Qi}\ \emph {et~al.}(2010)\citenamefont {Qi},
  \citenamefont {Hughes},\ and\ \citenamefont {Zhang}}]{PhysRevB.82.184516}%
  \BibitemOpen
  \bibfield  {author} {\bibinfo {author} {\bibfnamefont {X.-L.}\ \bibnamefont
  {Qi}}, \bibinfo {author} {\bibfnamefont {T.~L.}\ \bibnamefont {Hughes}}, \
  and\ \bibinfo {author} {\bibfnamefont {S.-C.}\ \bibnamefont {Zhang}},\ }\href
  {\doibase 10.1103/PhysRevB.82.184516} {\bibfield  {journal} {\bibinfo
  {journal} {Phys. Rev. B}\ }\textbf {\bibinfo {volume} {82}},\ \bibinfo
  {pages} {184516} (\bibinfo {year} {2010})}\BibitemShut {NoStop}%
\bibitem [{\citenamefont {Das}\ \emph {et~al.}(2012)\citenamefont {Das},
  \citenamefont {Ronen}, \citenamefont {Most}, \citenamefont {Oreg},
  \citenamefont {Heiblum},\ and\ \citenamefont {Shtrikman}}]{nphys2479}%
  \BibitemOpen
  \bibfield  {author} {\bibinfo {author} {\bibfnamefont {A.}~\bibnamefont
  {Das}}, \bibinfo {author} {\bibfnamefont {Y.}~\bibnamefont {Ronen}}, \bibinfo
  {author} {\bibfnamefont {Y.}~\bibnamefont {Most}}, \bibinfo {author}
  {\bibfnamefont {Y.}~\bibnamefont {Oreg}}, \bibinfo {author} {\bibfnamefont
  {M.}~\bibnamefont {Heiblum}}, \ and\ \bibinfo {author} {\bibfnamefont
  {H.}~\bibnamefont {Shtrikman}},\ }\href {\doibase 10.1038/nphys2479}
  {\bibfield  {journal} {\bibinfo  {journal} {Nature Physics}\ }\textbf
  {\bibinfo {volume} {8}},\ \bibinfo {pages} {887 EP } (\bibinfo {year}
  {2012})}\BibitemShut {NoStop}%
\bibitem [{\citenamefont {Fu}\ and\ \citenamefont
  {Berg}(2010)}]{PhysRevLett.105.097001}%
  \BibitemOpen
  \bibfield  {author} {\bibinfo {author} {\bibfnamefont {L.}~\bibnamefont
  {Fu}}\ and\ \bibinfo {author} {\bibfnamefont {E.}~\bibnamefont {Berg}},\
  }\href {\doibase 10.1103/PhysRevLett.105.097001} {\bibfield  {journal}
  {\bibinfo  {journal} {Phys. Rev. Lett.}\ }\textbf {\bibinfo {volume} {105}},\
  \bibinfo {pages} {097001} (\bibinfo {year} {2010})}\BibitemShut {NoStop}%
\bibitem [{\citenamefont {Haldane}(1988)}]{PhysRevLett.61.2015}%
  \BibitemOpen
  \bibfield  {author} {\bibinfo {author} {\bibfnamefont {F.~D.~M.}\
  \bibnamefont {Haldane}},\ }\href {\doibase 10.1103/PhysRevLett.61.2015}
  {\bibfield  {journal} {\bibinfo  {journal} {Phys. Rev. Lett.}\ }\textbf
  {\bibinfo {volume} {61}},\ \bibinfo {pages} {2015} (\bibinfo {year}
  {1988})}\BibitemShut {NoStop}%
\bibitem [{\citenamefont {Regnault}\ and\ \citenamefont
  {Bernevig}(2011)}]{PhysRevX.1.021014}%
  \BibitemOpen
  \bibfield  {author} {\bibinfo {author} {\bibfnamefont {N.}~\bibnamefont
  {Regnault}}\ and\ \bibinfo {author} {\bibfnamefont {B.~A.}\ \bibnamefont
  {Bernevig}},\ }\href {\doibase 10.1103/PhysRevX.1.021014} {\bibfield
  {journal} {\bibinfo  {journal} {Phys. Rev. X}\ }\textbf {\bibinfo {volume}
  {1}},\ \bibinfo {pages} {021014} (\bibinfo {year} {2011})}\BibitemShut
  {NoStop}%
\bibitem [{\citenamefont {Zhang}\ and\ \citenamefont
  {Qi}(2014)}]{PhysRevB.89.195144}%
  \BibitemOpen
  \bibfield  {author} {\bibinfo {author} {\bibfnamefont {Y.}~\bibnamefont
  {Zhang}}\ and\ \bibinfo {author} {\bibfnamefont {X.-L.}\ \bibnamefont {Qi}},\
  }\href {\doibase 10.1103/PhysRevB.89.195144} {\bibfield  {journal} {\bibinfo
  {journal} {Phys. Rev. B}\ }\textbf {\bibinfo {volume} {89}},\ \bibinfo
  {pages} {195144} (\bibinfo {year} {2014})}\BibitemShut {NoStop}%
\bibitem [{\citenamefont {Thonhauser}\ and\ \citenamefont
  {Vanderbilt}(2006)}]{PhysRevB.74.235111}%
  \BibitemOpen
  \bibfield  {author} {\bibinfo {author} {\bibfnamefont {T.}~\bibnamefont
  {Thonhauser}}\ and\ \bibinfo {author} {\bibfnamefont {D.}~\bibnamefont
  {Vanderbilt}},\ }\href {\doibase 10.1103/PhysRevB.74.235111} {\bibfield
  {journal} {\bibinfo  {journal} {Phys. Rev. B}\ }\textbf {\bibinfo {volume}
  {74}},\ \bibinfo {pages} {235111} (\bibinfo {year} {2006})}\BibitemShut
  {NoStop}%
\bibitem [{\citenamefont {Qi}\ and\ \citenamefont
  {Zhang}(2011)}]{RevModPhys.83.1057}%
  \BibitemOpen
  \bibfield  {author} {\bibinfo {author} {\bibfnamefont {X.-L.}\ \bibnamefont
  {Qi}}\ and\ \bibinfo {author} {\bibfnamefont {S.-C.}\ \bibnamefont {Zhang}},\
  }\href {\doibase 10.1103/RevModPhys.83.1057} {\bibfield  {journal} {\bibinfo
  {journal} {Rev. Mod. Phys.}\ }\textbf {\bibinfo {volume} {83}},\ \bibinfo
  {pages} {1057} (\bibinfo {year} {2011})}\BibitemShut {NoStop}%
\bibitem [{\citenamefont {Moore}\ and\ \citenamefont
  {Balents}(2007)}]{PhysRevB.75.121306}%
  \BibitemOpen
  \bibfield  {author} {\bibinfo {author} {\bibfnamefont {J.~E.}\ \bibnamefont
  {Moore}}\ and\ \bibinfo {author} {\bibfnamefont {L.}~\bibnamefont
  {Balents}},\ }\href {\doibase 10.1103/PhysRevB.75.121306} {\bibfield
  {journal} {\bibinfo  {journal} {Phys. Rev. B}\ }\textbf {\bibinfo {volume}
  {75}},\ \bibinfo {pages} {121306} (\bibinfo {year} {2007})}\BibitemShut
  {NoStop}%
\bibitem [{\citenamefont {Witten}(1983)}]{WITTEN1983422}%
  \BibitemOpen
  \bibfield  {author} {\bibinfo {author} {\bibfnamefont {E.}~\bibnamefont
  {Witten}},\ }\href {\doibase https://doi.org/10.1016/0550-3213(83)90063-9}
  {\bibfield  {journal} {\bibinfo  {journal} {Nuclear Physics B}\ }\textbf
  {\bibinfo {volume} {223}},\ \bibinfo {pages} {422 } (\bibinfo {year}
  {1983})}\BibitemShut {NoStop}%
\bibitem [{\citenamefont {Wang}\ \emph
  {et~al.}(2010{\natexlab{a}})\citenamefont {Wang}, \citenamefont {Qi},\ and\
  \citenamefont {Zhang}}]{PhysRevLett.105.256803}%
  \BibitemOpen
  \bibfield  {author} {\bibinfo {author} {\bibfnamefont {Z.}~\bibnamefont
  {Wang}}, \bibinfo {author} {\bibfnamefont {X.-L.}\ \bibnamefont {Qi}}, \ and\
  \bibinfo {author} {\bibfnamefont {S.-C.}\ \bibnamefont {Zhang}},\ }\href
  {\doibase 10.1103/PhysRevLett.105.256803} {\bibfield  {journal} {\bibinfo
  {journal} {Phys. Rev. Lett.}\ }\textbf {\bibinfo {volume} {105}},\ \bibinfo
  {pages} {256803} (\bibinfo {year} {2010}{\natexlab{a}})}\BibitemShut
  {NoStop}%
\bibitem [{\citenamefont {Wang}\ \emph
  {et~al.}(2010{\natexlab{b}})\citenamefont {Wang}, \citenamefont {Qi},\ and\
  \citenamefont {Zhang}}]{1367-2630-12-6-065007}%
  \BibitemOpen
  \bibfield  {author} {\bibinfo {author} {\bibfnamefont {Z.}~\bibnamefont
  {Wang}}, \bibinfo {author} {\bibfnamefont {X.-L.}\ \bibnamefont {Qi}}, \ and\
  \bibinfo {author} {\bibfnamefont {S.-C.}\ \bibnamefont {Zhang}},\ }\href
  {http://stacks.iop.org/1367-2630/12/i=6/a=065007} {\bibfield  {journal}
  {\bibinfo  {journal} {New Journal of Physics}\ }\textbf {\bibinfo {volume}
  {12}},\ \bibinfo {pages} {065007} (\bibinfo {year}
  {2010}{\natexlab{b}})}\BibitemShut {NoStop}%
\bibitem [{\citenamefont {Essin}\ and\ \citenamefont
  {Gurarie}(2011)}]{PhysRevB.84.125132}%
  \BibitemOpen
  \bibfield  {author} {\bibinfo {author} {\bibfnamefont {A.~M.}\ \bibnamefont
  {Essin}}\ and\ \bibinfo {author} {\bibfnamefont {V.}~\bibnamefont
  {Gurarie}},\ }\href {\doibase 10.1103/PhysRevB.84.125132} {\bibfield
  {journal} {\bibinfo  {journal} {Phys. Rev. B}\ }\textbf {\bibinfo {volume}
  {84}},\ \bibinfo {pages} {125132} (\bibinfo {year} {2011})}\BibitemShut
  {NoStop}%
\bibitem [{\citenamefont {Niu}\ \emph {et~al.}(1985)\citenamefont {Niu},
  \citenamefont {Thouless},\ and\ \citenamefont {Wu}}]{PhysRevB.31.3372}%
  \BibitemOpen
  \bibfield  {author} {\bibinfo {author} {\bibfnamefont {Q.}~\bibnamefont
  {Niu}}, \bibinfo {author} {\bibfnamefont {D.~J.}\ \bibnamefont {Thouless}}, \
  and\ \bibinfo {author} {\bibfnamefont {Y.-S.}\ \bibnamefont {Wu}},\ }\href
  {\doibase 10.1103/PhysRevB.31.3372} {\bibfield  {journal} {\bibinfo
  {journal} {Phys. Rev. B}\ }\textbf {\bibinfo {volume} {31}},\ \bibinfo
  {pages} {3372} (\bibinfo {year} {1985})}\BibitemShut {NoStop}%
\bibitem [{\citenamefont {Thouless}\ \emph {et~al.}(1982)\citenamefont
  {Thouless}, \citenamefont {Kohmoto}, \citenamefont {Nightingale},\ and\
  \citenamefont {den Nijs}}]{PhysRevLett.49.405}%
  \BibitemOpen
  \bibfield  {author} {\bibinfo {author} {\bibfnamefont {D.~J.}\ \bibnamefont
  {Thouless}}, \bibinfo {author} {\bibfnamefont {M.}~\bibnamefont {Kohmoto}},
  \bibinfo {author} {\bibfnamefont {M.~P.}\ \bibnamefont {Nightingale}}, \ and\
  \bibinfo {author} {\bibfnamefont {M.}~\bibnamefont {den Nijs}},\ }\href
  {\doibase 10.1103/PhysRevLett.49.405} {\bibfield  {journal} {\bibinfo
  {journal} {Phys. Rev. Lett.}\ }\textbf {\bibinfo {volume} {49}},\ \bibinfo
  {pages} {405} (\bibinfo {year} {1982})}\BibitemShut {NoStop}%
\bibitem [{\citenamefont {M.~V.~Berry}(1984)}]{Berry45}%
  \BibitemOpen
  \bibfield  {author} {\bibinfo {author} {\bibfnamefont {F.~R.~S.}\
  \bibnamefont {M.~V.~Berry}},\ }\href {\doibase 10.1098/rspa.1984.0023}
  {\bibfield  {journal} {\bibinfo  {journal} {Proceedings of the Royal Society
  of London A: Mathematical, Physical and Engineering Sciences}\ }\textbf
  {\bibinfo {volume} {392}},\ \bibinfo {pages} {45} (\bibinfo {year}
  {1984})}\BibitemShut {NoStop}%
\bibitem [{\citenamefont {Riwar}\ \emph {et~al.}(2016)\citenamefont {Riwar},
  \citenamefont {Houzet}, \citenamefont {Meyer},\ and\ \citenamefont
  {Nazarov}}]{ncomms11167}%
  \BibitemOpen
  \bibfield  {author} {\bibinfo {author} {\bibfnamefont {R.-P.}\ \bibnamefont
  {Riwar}}, \bibinfo {author} {\bibfnamefont {M.}~\bibnamefont {Houzet}},
  \bibinfo {author} {\bibfnamefont {J.~S.}\ \bibnamefont {Meyer}}, \ and\
  \bibinfo {author} {\bibfnamefont {Y.~V.}\ \bibnamefont {Nazarov}},\ }\href
  {\doibase 10.1038/ncomms11167} {\bibfield  {journal} {\bibinfo  {journal}
  {Nature Communications}\ }\textbf {\bibinfo {volume} {7}},\ \bibinfo {pages}
  {11167 EP } (\bibinfo {year} {2016})}\BibitemShut {NoStop}%
\bibitem [{\citenamefont {Andreev}(1964)}]{Andreev}%
  \BibitemOpen
  \bibfield  {author} {\bibinfo {author} {\bibfnamefont {A.}~\bibnamefont
  {Andreev}},\ }\href@noop {} {\bibfield  {journal} {\bibinfo  {journal} {Sov.
  Phys. JETP}\ }\textbf {\bibinfo {volume} {19}},\ \bibinfo {pages} {1228}
  (\bibinfo {year} {1964})}\BibitemShut {NoStop}%
\bibitem [{\citenamefont {de~Gennes}\ and\ \citenamefont
  {Saint-James}(1963)}]{DEGENNES1963151}%
  \BibitemOpen
  \bibfield  {author} {\bibinfo {author} {\bibfnamefont {P.}~\bibnamefont
  {de~Gennes}}\ and\ \bibinfo {author} {\bibfnamefont {D.}~\bibnamefont
  {Saint-James}},\ }\href {\doibase
  https://doi.org/10.1016/0031-9163(63)90148-3} {\bibfield  {journal} {\bibinfo
   {journal} {Physics Letters}\ }\textbf {\bibinfo {volume} {4}},\ \bibinfo
  {pages} {151 } (\bibinfo {year} {1963})}\BibitemShut {NoStop}%
\bibitem [{\citenamefont {Beenakker}\ and\ \citenamefont {van
  Houten}(1991)}]{PhysRevLett.66.3056}%
  \BibitemOpen
  \bibfield  {author} {\bibinfo {author} {\bibfnamefont {C.~W.~J.}\
  \bibnamefont {Beenakker}}\ and\ \bibinfo {author} {\bibfnamefont
  {H.}~\bibnamefont {van Houten}},\ }\href {\doibase
  10.1103/PhysRevLett.66.3056} {\bibfield  {journal} {\bibinfo  {journal}
  {Phys. Rev. Lett.}\ }\textbf {\bibinfo {volume} {66}},\ \bibinfo {pages}
  {3056} (\bibinfo {year} {1991})}\BibitemShut {NoStop}%
\bibitem [{\citenamefont {Kitaev}(2009)}]{doi:10.1063/1.3149495}%
  \BibitemOpen
  \bibfield  {author} {\bibinfo {author} {\bibfnamefont {A.}~\bibnamefont
  {Kitaev}},\ }\href {\doibase 10.1063/1.3149495} {\bibfield  {journal}
  {\bibinfo  {journal} {AIP Conference Proceedings}\ }\textbf {\bibinfo
  {volume} {1134}},\ \bibinfo {pages} {22} (\bibinfo {year}
  {2009})}\BibitemShut {NoStop}%
\bibitem [{\citenamefont {Lu}\ \emph {et~al.}(2015)\citenamefont {Lu},
  \citenamefont {Wang}, \citenamefont {Ye}, \citenamefont {Ran}, \citenamefont
  {Fu}, \citenamefont {Joannopoulos},\ and\ \citenamefont {Solja{\v
  c}i{\'c}}}]{Lu622}%
  \BibitemOpen
  \bibfield  {author} {\bibinfo {author} {\bibfnamefont {L.}~\bibnamefont
  {Lu}}, \bibinfo {author} {\bibfnamefont {Z.}~\bibnamefont {Wang}}, \bibinfo
  {author} {\bibfnamefont {D.}~\bibnamefont {Ye}}, \bibinfo {author}
  {\bibfnamefont {L.}~\bibnamefont {Ran}}, \bibinfo {author} {\bibfnamefont
  {L.}~\bibnamefont {Fu}}, \bibinfo {author} {\bibfnamefont {J.~D.}\
  \bibnamefont {Joannopoulos}}, \ and\ \bibinfo {author} {\bibfnamefont
  {M.}~\bibnamefont {Solja{\v c}i{\'c}}},\ }\href {\doibase
  10.1126/science.aaa9273} {\bibfield  {journal} {\bibinfo  {journal}
  {Science}\ }\textbf {\bibinfo {volume} {349}},\ \bibinfo {pages} {622}
  (\bibinfo {year} {2015})}\BibitemShut {NoStop}%
\bibitem [{\citenamefont {Soluyanov}\ \emph {et~al.}(2015)\citenamefont
  {Soluyanov}, \citenamefont {Gresch}, \citenamefont {Wang}, \citenamefont
  {Wu}, \citenamefont {Troyer}, \citenamefont {Dai},\ and\ \citenamefont
  {Bernevig}}]{nature15768}%
  \BibitemOpen
  \bibfield  {author} {\bibinfo {author} {\bibfnamefont {A.~A.}\ \bibnamefont
  {Soluyanov}}, \bibinfo {author} {\bibfnamefont {D.}~\bibnamefont {Gresch}},
  \bibinfo {author} {\bibfnamefont {Z.}~\bibnamefont {Wang}}, \bibinfo {author}
  {\bibfnamefont {Q.}~\bibnamefont {Wu}}, \bibinfo {author} {\bibfnamefont
  {M.}~\bibnamefont {Troyer}}, \bibinfo {author} {\bibfnamefont
  {X.}~\bibnamefont {Dai}}, \ and\ \bibinfo {author} {\bibfnamefont {B.~A.}\
  \bibnamefont {Bernevig}},\ }\href {http://dx.doi.org/10.1038/nature15768}
  {\bibfield  {journal} {\bibinfo  {journal} {Nature}\ }\textbf {\bibinfo
  {volume} {527}},\ \bibinfo {pages} {495 EP } (\bibinfo {year}
  {2015})}\BibitemShut {NoStop}%
\bibitem [{\citenamefont {Nazarov}\ and\ \citenamefont
  {Blanter}(2009)}]{Transport}%
  \BibitemOpen
  \bibfield  {author} {\bibinfo {author} {\bibfnamefont {Y.}~\bibnamefont
  {Nazarov}}\ and\ \bibinfo {author} {\bibfnamefont {Y.}~\bibnamefont
  {Blanter}},\ }\href@noop {} {\emph {\bibinfo {title} {Quantum Transport}}}\
  (\bibinfo  {publisher} {Cambridge University Press},\ \bibinfo {year}
  {2009})\BibitemShut {NoStop}%
\bibitem [{\citenamefont {Schön}\ and\ \citenamefont
  {Zaikin}(1990)}]{SCHON1990237}%
  \BibitemOpen
  \bibfield  {author} {\bibinfo {author} {\bibfnamefont {G.}~\bibnamefont
  {Schön}}\ and\ \bibinfo {author} {\bibfnamefont {A.}~\bibnamefont
  {Zaikin}},\ }\href {\doibase https://doi.org/10.1016/0370-1573(90)90156-V}
  {\bibfield  {journal} {\bibinfo  {journal} {Physics Reports}\ }\textbf
  {\bibinfo {volume} {198}},\ \bibinfo {pages} {237 } (\bibinfo {year}
  {1990})}\BibitemShut {NoStop}%
\bibitem [{\citenamefont {Beenakker}(1997)}]{RevModPhys.69.731}%
  \BibitemOpen
  \bibfield  {author} {\bibinfo {author} {\bibfnamefont {C.~W.~J.}\
  \bibnamefont {Beenakker}},\ }\href {\doibase 10.1103/RevModPhys.69.731}
  {\bibfield  {journal} {\bibinfo  {journal} {Rev. Mod. Phys.}\ }\textbf
  {\bibinfo {volume} {69}},\ \bibinfo {pages} {731} (\bibinfo {year}
  {1997})}\BibitemShut {NoStop}%
\bibitem [{\citenamefont {Eilenberger}(1968)}]{Eilenberger1968}%
  \BibitemOpen
  \bibfield  {author} {\bibinfo {author} {\bibfnamefont {G.}~\bibnamefont
  {Eilenberger}},\ }\href {\doibase 10.1007/BF01379803} {\bibfield  {journal}
  {\bibinfo  {journal} {Zeitschrift f{\"u}r Physik A Hadrons and nuclei}\
  }\textbf {\bibinfo {volume} {214}},\ \bibinfo {pages} {195} (\bibinfo {year}
  {1968})}\BibitemShut {NoStop}%
\bibitem [{\citenamefont {Qi}\ \emph {et~al.}(2008)\citenamefont {Qi},
  \citenamefont {Hughes},\ and\ \citenamefont {Zhang}}]{Zhang}%
  \BibitemOpen
  \bibfield  {author} {\bibinfo {author} {\bibfnamefont {X.}~\bibnamefont
  {Qi}}, \bibinfo {author} {\bibfnamefont {T.}~\bibnamefont {Hughes}}, \ and\
  \bibinfo {author} {\bibfnamefont {S.}~\bibnamefont {Zhang}},\ }\href
  {\doibase 10.1103/PhysRevB.78.195424} {\bibfield  {journal} {\bibinfo
  {journal} {Phys. Rev. B}\ }\textbf {\bibinfo {volume} {78}} (\bibinfo {year}
  {2008}),\ 10.1103/PhysRevB.78.195424}\BibitemShut {NoStop}%
\bibitem [{\citenamefont {Landauer}(1957)}]{5392683}%
  \BibitemOpen
  \bibfield  {author} {\bibinfo {author} {\bibfnamefont {R.}~\bibnamefont
  {Landauer}},\ }\href {\doibase 10.1147/rd.13.0223} {\bibfield  {journal}
  {\bibinfo  {journal} {IBM Journal of Research and Development}\ }\textbf
  {\bibinfo {volume} {1}},\ \bibinfo {pages} {223} (\bibinfo {year}
  {1957})}\BibitemShut {NoStop}%
\bibitem [{\citenamefont {Yokoyama}\ and\ \citenamefont
  {Nazarov}(2015)}]{PhysRevB.92.155437}%
  \BibitemOpen
  \bibfield  {author} {\bibinfo {author} {\bibfnamefont {T.}~\bibnamefont
  {Yokoyama}}\ and\ \bibinfo {author} {\bibfnamefont {Y.~V.}\ \bibnamefont
  {Nazarov}},\ }\href {\doibase 10.1103/PhysRevB.92.155437} {\bibfield
  {journal} {\bibinfo  {journal} {Phys. Rev. B}\ }\textbf {\bibinfo {volume}
  {92}},\ \bibinfo {pages} {155437} (\bibinfo {year} {2015})}\BibitemShut
  {NoStop}%
\bibitem [{\citenamefont {Golubov}\ \emph {et~al.}(2004)\citenamefont
  {Golubov}, \citenamefont {Kupriyanov},\ and\ \citenamefont
  {Il'ichev}}]{GolubovReview}%
  \BibitemOpen
  \bibfield  {author} {\bibinfo {author} {\bibfnamefont {A.~A.}\ \bibnamefont
  {Golubov}}, \bibinfo {author} {\bibfnamefont {M.~Y.}\ \bibnamefont
  {Kupriyanov}}, \ and\ \bibinfo {author} {\bibfnamefont {E.}~\bibnamefont
  {Il'ichev}},\ }\href {\doibase 10.1103/RevModPhys.76.411} {\bibfield
  {journal} {\bibinfo  {journal} {Rev. Mod. Phys.}\ }\textbf {\bibinfo {volume}
  {76}},\ \bibinfo {pages} {411} (\bibinfo {year} {2004})}\BibitemShut
  {NoStop}%
\bibitem [{\citenamefont {Nazarov}(2015)}]{Block}%
  \BibitemOpen
  \bibfield  {author} {\bibinfo {author} {\bibfnamefont {Y.~V.}\ \bibnamefont
  {Nazarov}},\ }\href {\doibase https://doi.org/10.1016/j.physe.2015.08.007}
  {\bibfield  {journal} {\bibinfo  {journal} {Physica E: Low-dimensional
  Systems and Nanostructures}\ }\textbf {\bibinfo {volume} {74}},\ \bibinfo
  {pages} {561 } (\bibinfo {year} {2015})}\BibitemShut {NoStop}%
\end{thebibliography}%

\end{document}